%% file: main.tex
\documentclass[twocolumn,prx,superscriptaddress]{revtex4-2}
\usepackage{amsmath,amssymb,amsthm,mathtools}
\usepackage{dsfont}
\usepackage{graphicx}
\usepackage{hyperref}

\newtheorem{theorem}{Theorem}[section]
\newtheorem{lemma}[theorem]{Lemma}
\newtheorem{proposition}[theorem]{Proposition}
\newtheorem{corollary}[theorem]{Corollary}
\newtheorem{conjecture}[theorem]{Conjecture}
\theoremstyle{definition}
\newtheorem{definition}[theorem]{Definition}

\theoremstyle{remark}
\newtheorem{remark}[theorem]{Remark}

\newcommand{\cT}{\mathcal{T}}
\newcommand{\cJ}{\mathcal{J}}
\newcommand{\cM}{\mathcal{M}}
\newcommand{\cB}{\mathcal{B}}
\newcommand{\cQ}{\mathcal{Q}}
\newcommand{\cV}{\mathcal{V}}

\newcommand{\cK}{\mathcal{K}}
\newcommand{\cI}{\mathcal{I}}
\newcommand{\cH}{\mathcal{H}}
\newcommand{\bbE}{\mathbb{E}}
\newcommand{\bbone}{\mathds{1}}
\newcommand{\tr}{\operatorname{tr}}
\newcommand{\spec}{\operatorname{spec}}
\newcommand{\spn}{\operatorname{span}}

\begin{document}

\title{Quantum Time Series Theory: Stationary Processes, Memory Spectra, and Measurement Records}
\author{Hai Wang}
\affiliation{School of Mathematics and Statistics, Nanjing University of Science and Technology, Nanjing, China}
\date{\today}

\begin{abstract}
Quantum experiments are naturally time series: every run is a sequence of interventions and outcomes, and the memory of an open system is written in its temporal statistics. However, the classical time-series framework cannot be translated directly: measuring a process and forgetting the outcome differs from not measuring at all, so no single family of joint probability distributions describes a quantum process. In this work, we provide one theory for stationary quantum time series based on the process-tensor framework, where structural theorems are established for the model class. We construct a finite-dimensional causal transfer map for stationary combs of finite quantum Markov order relative to a fixed informationally complete causal break, and we prove spectral criteria for ergodicity and mixing of the process and of its measurement records. In this new framework, we identify a no-broadcasting obstruction, a time-series counterpart of the no-broadcasting theorem of Ref.~\cite{Barnum1996}: a perfectly preserved noncommutative memory mode is invisible to the record generated by any fixed instrument whose dynamics preserves that block, an obstruction with no classical counterpart: classical content can be read out without disturbing it, whereas reading noncommuting quantum content necessarily disturbs it, so perfect preservation and record visibility cannot coexist. This obstruction reveals an essential difference between quantum and classical time series: what a process remembers and what an observer can read out are not the same, and it is precisely where quantum memory can hide from observation. We believe that these results supply the stationary foundation on which quantum spectral analysis, model identification, and single-trajectory inference can be built.
\end{abstract}

\maketitle

\section{Introduction}

Time series analysis is the statistical theory of ordered data, and it provides one of the most successful examples of structure learned from a single observed trajectory \cite{BrockwellDavis1991,Hamilton1994,ShumwayStoffer2017}. A stationary process is one whose statistical law is invariant under time translation. The Kolmogorov extension theorem guarantees that a consistent family of finite-dimensional distributions uniquely defines a single process \cite{Kolmogorov1950}. Birkhoff's ergodic theorem equates time averages along one trajectory with ensemble averages \cite{Birkhoff1931}, making single-trajectory inference possible. The Wiener--Khinchin spectral representation turns temporal correlation into a frequency-domain object \cite{Wiener1949,Khinchin1934}. Finite-memory parametrized families, such as Markov chains and ARMA models, make the theory estimable in practice \cite{BoxJenkins1970}. Together, these results converted temporal dependence from a nuisance into structure: forecasting, system identification, signal processing, and control all rest on them. The conceptual lesson is that the way a process remembers and forgets is itself a learnable, classifiable object.

On the other hand, quantum information theory is built on dynamical primitives. A quantum state is the quantum analogue of a random variable, and a quantum channel that of a stochastic map \cite{NielsenChuang2010}. The temporal evolution of a quantum system, especially an open one, is generically non-Markovian: the environment remembers. Continuous monitoring of quantum systems produces measurement records whose temporal correlations carry information about the dynamics \cite{Belavkin1992,KM2003,KM2004,Benoist2023}, and every quantum experiment, from benchmarking to sensing, is literally a time series of interventions and outcomes. Several pieces of a quantum theory of temporal structure are already known: the process tensor characterizes non-Markovian multi-time processes \cite{Pollock2018operational,Pollock2018complete,Milz2021}, quantum trajectory theory analyzes continuously measured systems \cite{KM2003,KM2004,Benoist2023}, quantum Markov chains have their own statistical theory \cite{vanHorssen2015,Guta2015}, and the algebraic quantum probability tradition defines stationary quantum processes abstractly \cite{Accardi1982,Kummerer1985}.

However, none of these pieces supplies the operational time-series level of organization: a theory whose objects are stationary processes, whose questions concern ergodicity, spectra, and finite-memory models, and in which all of the above pieces sit as special cases. Such a theory would unify the description of open-system memory, monitored dynamics, and device-level temporal correlations in the way classical time series analysis unifies forecasting and identification.

The obstruction to such a theory is foundational. The Kolmogorov consistency condition, marginalization, reads ``measure and forget the outcome''. Quantum-mechanically, this differs from ``not measuring at all'', because measurement disturbs the system. Consequently, no single family of joint probability distributions describes a quantum process: outcome statistics depend on which instruments were applied, and incompatible choices cannot be embedded in one distribution. Two correlates follow. First, there is no canonical two-time correlation: what happens between the two times, idle evolution or measurement, changes the answer, so the classical autocorrelation function has no unique quantum descendant. Second, ergodicity becomes subtle: if observation backacts on the process, the sense in which a single trajectory represents the ensemble needs re-examination. Any quantum time series theory must solve these problems before it can formulate the questions that classical time-series analysis answers.

The framework we build on has two ancestries. On the operational side, quantum combs were introduced by Chiribella, D'Ariano, and Perinotti \cite{CDP2008,CDP2009} and were developed into a complete account of non-Markovian multi-time processes by Pollock et al.\ \cite{Pollock2018operational,Pollock2018complete} (reviewed in \cite{Milz2021,Keeling2026}). The existence of infinite processes rests on the generalized extension theorem (GET) of Milz et al.\ \cite{Milz2020}, which plays the role of the Kolmogorov extension theorem in our construction. The notion of memory we use is the quantum Markov order of Taranto et al.\ \cite{Taranto2019markovorder,Taranto2019structure}, whose recoverability aspects were studied in \cite{Taranto2021}, and combs as higher-order operations are treated in \cite{Taranto2025}. Time-translation-invariant process tensors have appeared in the numerical literature \cite{Link2024,Garbellini2026}, where translation invariance provides an ansatz that accelerates computations. Our work is complementary to these studies, in that stationarity is imposed as an operational symmetry on the comb family itself.

On the mathematical side, our transfer map parallels the finitely correlated states of Fannes, Nachtergaele, and Werner \cite{FNW1992}, constructed here for interventions rather than states, and memory channels \cite{Kretschmann2005} provide a complementary input-output picture. Hidden quantum Markov models \cite{Monras2011,MonrasWinter2016} and matrix-product approaches to stochastic modelling \cite{Yang2018} correspond to the single-instrument, order-one sector of what we call the record layer. The difference is one of domain: these constructions attach a finite-dimensional memory to a state or to a single fixed instrument. In contrast, our transfer map resolves every intervention element on a stationary comb, which is what our spectral criteria and the record ergodic theorem require.

On the ergodic-theory side, the study of quantum measurement records was pioneered by Belavkin \cite{Belavkin1992} and by K\"ummerer and Maassen \cite{KM2003,KM2004} and continues through \cite{vanHorssen2015,Guta2015,Benoist2023,Benoist2025}, all in settings where the environment carries no process-tensor memory. For collision models specifically, the repeated-interaction literature established approach-to-stationarity and return-to-equilibrium results \cite{BruneauJoyeMerkli2006} on the same physical objects. Those results concern state-level relaxation in concrete Hamiltonian models, while our spectral criteria here concern the operational comb. In terms of stationary quantum processes, Accardi, Frigerio, and Lewis and K\"ummerer first defined them in a non-operational way \cite{Accardi1982,Kummerer1985}. Our approach is complementary in that every object is defined directly in terms of interventions. Markovianity provides a separate axis of comparison, distinct from this difference in how stationary processes are defined. Closest in spirit to our two-level structure, Taranto and Milz showed that records Markovian under every probing can still require memory in the process \cite{TarantoMilz2023}, and Milz et al.\ studied when the process layer admits a classical description at all \cite{Milz2020classical}. Both separations concern Markovianity, while ours concern mixing rates and spectra. The increment of the present work is to organize stationarity, finite-order process memory, and operator-spectral criteria within one operational comb framework, and to identify no broadcasting as a quantum-specific source of record invisibility.

In this paper, we build the theory on the process-tensor foundation and prove structural theorems for the finite-order model class specified below. The basic object is a family of quantum combs, marginalization is replaced by identity insertion (do nothing), the causally correct neutral operation \cite{Pollock2018operational,Milz2020}, and the GET then provides the existence of infinite processes. On this foundation, the classical questions can be posed operationally. We answer them for the simplest nontrivial model class, stationary processes with finite quantum Markov order relative to a fixed informationally complete (IC) causal break \cite{Taranto2019markovorder,Taranto2019structure}. Concretely, we address the following questions:
\begin{enumerate}
\item[(Q1)] What is a \emph{stationary} quantum process, operationally? (Sec.~\ref{sec:stationary})
\item[(Q2)] Does a stationary quantum process with finite quantum Markov order admit a finite-dimensional \emph{transfer map} representation, analogous to the finitely correlated states of Fannes, Nachtergaele, and Werner \cite{FNW1992}? (Sec.~\ref{sec:transfer})
\item[(Q3)] What are the spectral criteria for ergodicity and mixing, and what replaces the autocorrelation function? (Sec.~\ref{sec:spectral})
\item[(Q4)] When can a single measurement trajectory be used to infer the process. In other words, is there a quantum ergodic theorem with rates? (Sec.~\ref{sec:spectral}, Theorem~\ref{thm:record})
\end{enumerate}
We give definite answers to all four questions: the answers include an algebraic signature of measurement backaction, a two-level splitting of the autocorrelation function, and an independence of the two levels of mixing, whose quantum-specific obstruction is captured by the record-invisibility theorem (Theorem~\ref{thm:nbl}, proved in Appendix~\ref{app:nbl}). The two-level independence itself is a general structural feature shared by classical and quantum models. No broadcasting supplies the obstruction that has no classical counterpart. The answers are summarized in prose in Sec.~\ref{sec:summary}, with pointers to the precise statements.

This paper is organized as follows. In Sec.~\ref{sec:summary}, we summarize the main results in prose, with pointers to the precise statements and proofs. In Sec.~\ref{sec:prelim}, we fix notation and recall quantum combs and Markov order. In Sec.~\ref{sec:stationary}, we define stationary quantum time series and prove their unique covariant extension. In Sec.~\ref{sec:transfer}, we construct the causal transfer map. In Sec.~\ref{sec:spectral}, we develop the spectral theory: the two autocorrelations, the ergodicity and mixing criteria, and the record ergodic theorem. Section~\ref{sec:scenarios} collects physical readings of the theory, Sec.~\ref{sec:numerics} reports the collision-model study, and Sec.~\ref{sec:discussion} collects open problems. Appendices~\ref{app:stationary}--\ref{app:nbl} give the proofs and the numerical details, each headed by the main-text statement it supplies.

\noindent\textbf{Scope and assumptions.} Four levels of restriction must be kept distinct. (i) The finite-dimensional representation assumes finite quantum Markov order $\ell$. (ii) That Markov order is relative to one fixed, full-rank IC causal break. It is not a statement about every possible probing instrument. (iii) The matrix order, order-unit property, and complete positivity are established on the operational cone generated by measure-and-prepare interventions and identity insertions, although the functional-space dimension bound and the generating formula are defined for arbitrary intervention elements. (iv) The extension of the operational order to coherent interventions is an open problem, formulated below as Conjecture~\ref{conj:coherent}. The spectral criteria for intrinsic process memory use the idle transfer map. The record limit theorems additionally concern a fixed instrument and its instrument-dressed map.

\section{Summary of results}
\label{sec:summary}

In this section, we provide an informal summary of our main results for the model class studied in this paper: stationary quantum time series of finite quantum Markov order relative to a fixed IC causal break, the simplest nontrivial model class of the theory. The mathematically rigorous formulation of all statements, together with their proofs, will be given in the subsequent sections.

\noindent\textbf{Main result 1 (stationary quantum time series exist: Definition~\ref{def:stationary} and Proposition~\ref{prop:extension}).} The basic object is a family of quantum combs, one per finite time window, consistent under identity insertion and covariant under the time shift (Definition~\ref{def:stationary}). Every such family extends uniquely to a two-sided infinite process (Proposition~\ref{prop:extension}, proved in Appendix~\ref{app:stationary}). This is the operational replacement for the Kolmogorov extension theorem, whose classical consistency condition fails quantumly. In other words, it is what makes the classical questions about stationary processes well posed for quantum experiments.

\noindent\textbf{Main result 2 (finite Markov order means finite-dimensional memory: Proposition~\ref{prop:triple}).} Every stationary quantum time series of Markov order $\ell$ relative to the fixed informationally complete causal break admits a generating triple $(\cB,\{\bbE_{\hat\xi}\},\rho)$ with $\dim\cB\le d^{4\ell}$ (Proposition~\ref{prop:triple}), where $d$ is the system Hilbert-space dimension, or an upper bound on it if the local spaces vary: all multi-time statistics factor through a finite-dimensional memory, $\cT_{[1,n]}[\xi_1\otimes\cdots\otimes\xi_n]=\rho\big(\bbE_{\hat\xi_n}\circ\cdots\circ\bbE_{\hat\xi_1}(\bbone)\big)$, with one update map per intervention element. The memory space is intrinsic in the sense that it is identified with the span of all conditional future functionals. Its order-unit and complete-positivity structure is established on the operational cone generated by measure-and-prepare interventions and identity insertions (Lemma~\ref{lem:order}). The extension of the operational order from the measure-and-prepare/identity operational cone to the full cone of coherent interventions remains conjectural (Conjecture~\ref{conj:coherent}). Within this structure, the identity intervention is the unique intervention whose transfer map is unital for every process in the class (Remark~\ref{rem:unital}). The normalization axiom $\rho\circ\bbE_{\hat{\cM}}=\rho$ is exactly the causality of the comb.

\noindent\textbf{Main result 3 (the peripheral spectrum is the intrinsic process memory: Theorems~\ref{thm:ergodic}--\ref{thm:converse}).} The autocorrelation splits into an intrinsic process-memory correlation across idle gaps, governed by the idle transfer map $\bbE_\cI$, and an instrument-specific record correlation, governed by the instrument-dressed map $\bbE_\cJ$ (Definition~\ref{def:idlegap}). For the idle dynamics, ergodicity holds if and only if the eigenvalue $1$ of $\bbE_\cI$ is simple (Theorem~\ref{thm:ergodic}), and exponential mixing holds if and only if the peripheral spectrum is trivial. The intrinsic process-memory rate is set by the subleading eigenvalue modulus of $\bbE_\cI$, up to polynomial factors from Jordan blocks (Theorems \ref{thm:gap} and \ref{thm:converse}, proved in Appendix~\ref{app:spectral}).

\noindent\textbf{Main result 4 (single-trajectory inference at the record level: Theorem~\ref{thm:record}).} Fix a finite-outcome measure-and-prepare instrument and a bounded real-valued function of its record. If the instrument-dressed map $\bbE_\cJ$ has a spectral gap, the measurement record is exponentially $\alpha$-mixing and satisfies separately stated law-of-large-numbers and central-limit conclusions, with asymptotic variance $\sigma^2=\gamma_0+2\sum_{k\ge1}\gamma_k$ and the degenerate case $\sigma^2=0$ treated explicitly (Theorem~\ref{thm:record}). The resulting single-trajectory inference rate is instrument-specific: it is controlled by the spectrum of $\bbE_\cJ$, not by the intrinsic idle-memory spectrum of $\bbE_\cI$.

\noindent\textbf{Main result 5 (finite-dimensional consistency checks: Sec.~\ref{sec:numerics}).} Analytic chain formulas and double-precision diagonalization on qubit collision models of Markov order $2$ check model-specific record decay rates to five to six decimal places (Sec.~\ref{sec:numerics}, with formulas and sanity checks in Appendix~\ref{app:numerics}). They are not independent validations of the general theorems. The examples also display the two-level asymmetry in both directions: a record can be i.i.d.\ while the idle process fails to mix, and a key-locked classical memory can keep a record from mixing while the idle process mixes.

The table below situates the framework relative to classical time-series concepts and collects the scope qualifications stated above.

\begin{table*}[t]
\centering
\small
\begin{tabular}{p{3.9cm} p{6.4cm} p{3.1cm}}
\hline
\textbf{Classical time series} & \textbf{Quantum time series (this work)} & \textbf{Classical counterpart?} \\\hline
Joint distributions + Kolmogorov consistency & Comb family + identity-insertion consistency (GET) & Fails quantumly \\
Autocorrelation function & Two levels: idle-gap (process level, $\bbE_{\cI}$) and record level ($\bbE_{\cJ}$) & Splits quantumly \\
Transition/transfer structure & Causal transfer map. Identity insertion is the unique intervention unital for every process in the class & Partial \\
Markov order $k$ & Instrument-relative quantum Markov order $\ell$. Memory dimension $\le d^{4\ell}$ & Generalizes \\
Ergodicity and mixing & Spectral criteria on $\bbE_{\cI}$. Mixing at the two levels is independent (both implications fail) & Classical and quantum \\
--- & Coherent-intervention obstruction (temporal interference, Conjecture~\ref{conj:coherent}) & No counterpart \\\hline
\end{tabular}
\end{table*}

\section{Preliminaries}
\label{sec:prelim}

In this section, we fix notation and recall the two ingredients on which the theory rests: quantum combs, which play the role of joint distributions, and quantum Markov order, which plays the role of finite memory. Readers already fluent in the process-tensor formalism may proceed to Sec.~\ref{sec:stationary}.

\subsection{Process tensors and combs}

For notational simplicity, we assume that all local Hilbert spaces have the same dimension $d$. If they vary, $d$ denotes a uniform upper bound on the local system dimensions.

At each time $t$ the experimenter may apply an \emph{instrument} $\cJ_t=\{\hat{\cM}_{i_t}\}$: a collection of trace-nonincreasing CP maps $\hat{\cM}_{i_t}:\cB_1(\cH_t^{\mathtt{i}})\to\cB_1(\cH_t^{\mathtt{o}})$ summing to a channel \cite{DaviesLewis1970,Ozawa1984}. We work in the Choi picture \cite{Choi1975,Jamiolkowski1972}: an intervention element is an operator $\xi\in\cV_t:=\cB(\cH_t^{\mathtt{o}}\otimes\cH_t^{\mathtt{i}})$, and $\xi\ge0$ iff it is the Choi operator of a CP map. A $k$-step process is a comb $\cT_{\Lambda_k}$, a multilinear functional on tuples of intervention elements, normalized to $1$ on all tuples of deterministic operations and nonnegative on tester elements \cite{CDP2008,CDP2009,Milz2020}. Equivalently, a positive Choi operator $\Upsilon_{1:k}$ satisfying the causality hierarchy
\begin{equation}
\tr_{\mathtt{o}_k}\Upsilon_{1:k}=\bbone_{\mathtt{i}_k}\otimes\Upsilon_{1:k-1},\qquad \tr\Upsilon_{1:k}=d^{\,k}.
\end{equation}
Outcome probabilities are given by the link product, $p(i_1,\dots,i_k\,|\,\cJ_1,\dots,\cJ_k)=\tr\big[\Upsilon_{1:k}\,(\cM_{i_k}\otimes\cdots\otimes\cM_{i_1})^{T}\big]$. Here and throughout, $\cM$ denotes the Choi operator of the map $\hat{\cM}$, not the map itself, and $T$ denotes transposition on $\cB(\cH^{\mathtt{o}}\otimes\cH^{\mathtt{i}})$ relative to the product basis of the output and input Hilbert spaces, applied slotwise to tensor products. The \emph{identity intervention} $\cI$ (the identity channel, ``do nothing'') plays a distinguished role: marginalization is defined by inserting $\cI$, not by measuring and forgetting. It is the neutral marginalizing operation in the comb, not a suspension of physical evolution: the system continues to evolve between slots, and inserting $\cI$ merely leaves the slot unprobed. This distinction is the operational root of the failure of the classical Kolmogorov consistency condition \cite{Milz2020}.

The link product admits two readings worth keeping in mind. For a memoryless (Markov) process, $\Upsilon_{1:k}$ factorizes into a product of channel Choi operators, and the link product reduces to the Born rule chained through the channels. Memory is exactly the failure of this factorization \cite{Pollock2018complete}. The causality hierarchy, in turn, is the statement that later interventions cannot signal to earlier statistics: it is what makes $\cT$ a \emph{process} rather than an arbitrary multilinear form, and it is the axiom that will reappear as the normalization condition of the transfer map in Sec.~\ref{sec:transfer}.

We write $\cV_\Lambda=\bigotimes_{t\in\Lambda}\cV_t$ for finite intervals $\Lambda\Subset\mathbb{Z}$, $\mathfrak{V}=\overline{\bigcup_\Lambda\cV_\Lambda}$ for the quasilocal algebra of intervention operators, and $\sigma$ for the shift automorphism.

\subsection{Quantum Markov order}

Classical finite memory is counted in steps: a Markov chain of order $\ell$ forgets everything beyond the last $\ell$ symbols. The quantum analogue, introduced by Taranto et al.\ \cite{Taranto2019markovorder,Taranto2019structure}, is formulated through instruments rather than states, because ``what the process remembers'' depends on how the intervening slots are probed. The precise definition is the following:

\begin{definition}[{Quantum Markov order \cite{Taranto2019markovorder,Taranto2019structure}}]
A process has Markov order $\ell$ relative to an instrument sequence $\cJ_M$ acting on a memory block $M$ of $\ell$ consecutive steps if, for every outcome $x_M$ with operation $O_M^{(x_M)}$,
\begin{equation}\label{eq:markovorder}
\tr_{M}\big[O_M^{(x_M)}\,\Upsilon_{FMH}\big]=\Upsilon_{F}^{(x_M)}\otimes\Upsilon_{H}^{(x_M)},
\end{equation}
i.e., conditioning on each break outcome factorizes the conditional future from the conditional history.
\end{definition}

The physical content is best read through the \emph{causal break}: an agent who measures the middle block with an informationally complete (IC) POVM and reprepares it from an IC set erases the bridge between past and future. Equation \eqref{eq:markovorder} says that after such an erasure, the future depends on the history only through the break outcome. Two caveats matter here. First, Markov order is instrument-relative by construction. Taranto et al.\ showed that no non-Markovian process has finite order relative to \emph{all} instruments \cite{Taranto2019markovorder}, so ``finite memory'' is always relative to a probing class, and we fix an IC causal break throughout. Second, the factorization \eqref{eq:markovorder} is an identity of operators, not only of probabilities: it constrains the response to arbitrary further interventions on either side.

Two facts from \cite{Taranto2019structure} will be used. First, for an \emph{informationally complete} (IC) history-blocking instrument one has the block expansion
\begin{equation}\label{eq:blockexpansion}
\Upsilon_{FMH}=\sum_{x}\Upsilon_{F}^{(x)}\otimes\Delta_{M}^{(x)}\otimes\Upsilon_{H}^{(x)},
\end{equation}
where $\{\Delta_M^{(x)}\}$ is a dual set to the instrument Choi operators (not individually positive in general). The index $x$ labels a joint measure-and-prepare outcome of the causal break on $M$. And if a basis subfamily is selected below, the instrument operators, the dual operators, and the sum all use that same label set. All contractions use the input/output-leg ordering and Choi transpose of the link product above, while $\Upsilon_F^{(x)}$ and $\Upsilon_H^{(x)}$ carry the remaining future and history legs. Second, for a fixed system Hilbert-space dimension $d$, the operator space of the $\ell$-step memory block is
\begin{equation}\label{eq:memorydim}
W:=\cB\big(\cH_{k-1}^{\mathtt{o}}\otimes\cdots\otimes\cH_{k-\ell}^{\mathtt{i}}\big),\qquad \dim W=d^{4\ell}.
\end{equation}
A \emph{causal break} \cite{Pollock2018complete} (measure with an IC POVM and independently reprepare from an IC set at every step) is IC, so \eqref{eq:blockexpansion} applies to it. We take the IC POVM elements $\Pi^{(a)}$ to be full rank \cite{Prugovecki1977,Renes2004}. For example, for a qubit, let $\Pi_a^{\mathrm{SIC}}$ denote the elements of the tetrahedral SIC POVM. The depolarized elements $(1-\epsilon)\Pi_a^{\mathrm{SIC}}+\epsilon\,\bbone/4$, with $\epsilon\in(0,1)$, define a full-rank IC POVM. Full rank will provide a uniform lower bound on record probabilities (Lemma~\ref{lem:order}).

\subsection{Notation and terminology}
\label{sec:notation}

Table~\ref{tab:notation} collects the symbols used in the main statements. Throughout, \emph{identity insertion} names the neutral operation $\cI$ (informally, ``do nothing''), \emph{idle dynamics} names the transfer dynamics generated by $\bbE_{\cI}$, and \emph{idle-gap correlation} names the corresponding process-level correlator. We use these terms consistently rather than treating measurement followed by forgetting as identity insertion.

\begin{table*}[t]
\centering
\small
\caption{Notation and the two rate levels.}\label{tab:notation}
\begin{tabular}{p{3.0cm} p{10.2cm}}
\hline
\textbf{Symbol/term} & \textbf{Meaning} \\\hline
$\cT_\Lambda$ & Operational comb of the process on the finite time window $\Lambda$. \\
$\bbE_{\hat\xi}$ & Linear causal transfer map associated with an intervention element $\xi$. \\
$\bbE_\cI$ & Idle transfer map for identity insertions. Its spectrum controls intrinsic process-memory ergodicity and decay rates. \\
$\bbE_\cJ$ & Instrument-dressed record map $\sum_i\bbE_{\hat{\cM}_i}$ for a fixed instrument $\cJ$. Its spectrum controls instrument-specific record rates. \\
$\rho$ & Boundary state/readout functional of the generating triple, normalized by $\rho\circ\bbE_{\hat{\cM}}=\rho$ for channels $\hat{\cM}$. \\
$\cB$ & Abstract finite-dimensional vector space of intervention functionals in a generating triple. In the construction it is realized by $\cQ$, not by a physical memory Hilbert space. \\
$\cQ$ & Concrete span of conditional future functionals. Its operational positive cone is defined by tests on tuples of measure-and-prepare interventions and identity insertions. The cone is a subset of this vector space, not a Hilbert space. \\
$\cH_t^{\mathtt{i}},\cH_t^{\mathtt{o}}$ & Physical input/output Hilbert spaces at time $t$, distinct from the abstract functional spaces $\cB$ and $\cQ$. \\
$\ell$ & Quantum Markov order relative to the fixed full-rank IC causal break. \\
$d$ & System Hilbert-space dimension, or a uniform upper bound on local system dimensions when they vary. \\
$C_{\xi_A,\xi_B}(\tau)$, $\Delta(\tau)$ & General idle-gap correlation and its collision-model specialization, respectively. Both are intrinsic process-memory correlators governed by $\bbE_\cI$ (or its model realization $N$). \\
Record ACF, $C(\tau)$ & Autocorrelation of a classical outcome record for a fixed instrument. An instrument-specific correlator governed by $\bbE_\cJ$ (or its model realization $M$). \\\hline
\end{tabular}
\end{table*}

Unhatted symbols denote Choi operators (operators on $\cH^{\mathtt{o}}\otimes\cH^{\mathtt{i}}$), while hatted symbols denote the corresponding maps or interventions. Thus $\xi$ and $\Upsilon$ are Choi operators, $\hat\xi$ is the map corresponding to $\xi$, and $\cM$ and $\hat{\cM}$ are respectively the Choi operator and CP map of an instrument element. Transfer maps are indexed by the corresponding intervention maps, as in $\bbE_{\hat\xi}$ and $\bbE_{\hat{\cM}}$.

\section{Stationary quantum time series}
\label{sec:stationary}

In this section, we define the central object of the paper and establish its well-definedness on the two-sided infinite time axis. The definition translates the classical notion of a stationary process, replacing joint distributions by combs and marginalization by the identity insertion:

\begin{definition}[Stationary quantum time series]\label{def:stationary}
A family $\{\cT_\Lambda\}$ of combs on finite subsets $\Lambda\subset\mathbb{Z}$ (intervals and finite unions of intervals alike) is \emph{stationary} if
\begin{itemize}
\item[(i)] (consistency) for $\Lambda_k\subseteq\Lambda_\ell$, $\cT_{\Lambda_k}=\cT_{\Lambda_\ell}^{|\Lambda_k}$, where marginalization inserts the identity intervention at the dropped slots, whether or not they lie at the boundary of an interval.
\item[(ii)] (translation covariance) $\cT_{\Lambda+1}[\sigma(\xi_\Lambda)]=\cT_{\Lambda}[\xi_\Lambda]$ for all $\xi_\Lambda\in\cV_\Lambda$.
\end{itemize}
\end{definition}

Here $\cT_{\Lambda_\ell}^{|\Lambda_k}$ denotes the restriction of $\cT_{\Lambda_\ell}$ to the time slots in $\Lambda_k$, obtained by identity insertion on $\Lambda_\ell\setminus\Lambda_k$. The map $\sigma$ is the one-step time-shift automorphism on $\mathfrak{V}$, with $\sigma(\cV_\Lambda)=\cV_{\Lambda+1}$ and $\Lambda+1=\{t+1:t\in\Lambda\}$.

Two remarks bear on this definition. Condition (i) is the only causally available notion of forgetting: any operation that reads the dropped slots, even with its outcome discarded, would in general change the statistics of the remaining slots, so the identity insertion is the unique neutral marginalization. Condition (ii) corresponds to the shift invariance condition in the classical time series theory, now required on every instrument choice at once. It may be alarming that consistency under identity insertion is too weak to pin down a process, since it never tests the dropped slots. The GET of Milz et al.~\cite{Milz2020} guarantees this is not so: consistency on the operational domain is exactly what is needed for a unique extension:

\begin{proposition}[Unique covariant extension]\label{prop:extension}
A stationary comb family extends uniquely to a comb $\overline{\cT}^\sharp$ on the quasilocal domain $\overline{\cK}^\sharp_{\mathbb{Z}}$ of the GET, and the extension is translation-covariant: $\overline{\cT}^\sharp[\sigma(\xi)]=\overline{\cT}^\sharp[\xi]$.
\end{proposition}

\begin{proof}[Proof sketch]
Existence and uniqueness of $\overline{\cT}^\sharp$ follow from the GET \cite{Milz2020}. The shifted functional $\cT'[\xi]:=\overline{\cT}^\sharp[\sigma(\xi)]$ is well defined because the shift preserves finite-interval liftings and their causal compatibility, and its finite marginals coincide with those of $\overline{\cT}^\sharp$ by (ii). Uniqueness forces $\cT'=\overline{\cT}^\sharp$.
\end{proof}

For stationary quantum time series, this proposition plays the role of the Kolmogorov extension theorem: the bi-infinite stationary process exists and is unique as an operational object. The GET supplies existence and uniqueness, while condition (ii) in Definition~\ref{def:stationary} supplies translation covariance. Note that uniqueness holds for the comb, not for any underlying system--environment dilation, of which there are many. Every representation we build below is a representation of the comb itself.

We calibrate the definition on two anchor examples. A classical stationary process embeds diagonally: its combs are the classical joint distributions, identity insertion is marginalization, and the two consistency notions coincide. At the opposite extreme, the process that applies a fixed unitary to the system at every step and nothing else is stationary and remembers everything. Its correlations never decay, and it will reappear as the boundary case of the spectral criteria in Sec.~\ref{sec:spectral}.

A fixed instrument $\cJ$ repeated at every step produces a classical outcome record. Stationarity of the process implies stationarity of the record. The converse holds if the instrument family is tomographically complete \cite{White2022tomography}.

\section{Causal transfer maps}
\label{sec:transfer}

A classical order-$\ell$ Markov chain is summarized by finitely many numbers: the transition rule on the last $\ell$ symbols. We now construct the quantum analogue: a finite-dimensional operator space that plays the role of the memory, a family of maps on it that play the role of transition matrices, one per intervention, and a state that reads the statistics out. The construction is intrinsic: the memory is not a guessed ancillary system but the space of operationally distinguishable futures.

\subsection{The generating triple}

Two positive cones are distinguished throughout. The intervention cone $\mathcal O\subset\cV$ is the measure-and-prepare cone generated by positive combinations of causal-break Choi elements and identity insertions. The functional cone $\cQ_+\subset\cQ$ is defined below by evaluations on the operational test tuples $\mathfrak{T}_{\mathrm{op}}$. Its elements are conditional future functionals, not intervention operators.

The representation is packaged in the following definition:

\begin{definition}[Causal transfer map and generating triple]\label{def:triple}
A \emph{generating triple} $(\cB,\{\bbE_{\hat\xi}\},\rho)$ for a stationary comb family consists of: a finite-dimensional matrix-ordered space $\cB$ with order unit $\bbone$ (the matrix order used throughout this paper is the operational one constructed in Lemma~\ref{lem:order}). For every $\xi\in\cV$ a linear map $\bbE_{\hat\xi}:\cB\to\cB$ with
\begin{itemize}
\item (complete positivity) $\bbE_{\hat\xi}$ is CP whenever $\xi\ge0$ lies in $\mathcal O$. Whether this extends to coherent intervention elements is part of Conjecture~\ref{conj:coherent}.
\item (identity-insertion unitality) $\bbE_{\cI}(\bbone)=\bbone$.
\item (linearity) $\xi\mapsto\bbE_{\hat\xi}$ is linear.
\end{itemize}
and a state $\rho$ on $\cB$ satisfying the \emph{normalization condition} $\rho\circ\bbE_{\hat{\cM}}=\rho$ for every channel $\hat{\cM}$. The triple \emph{generates} the family if, for all finite windows,
\begin{equation}\label{eq:generating}
\cT_{[1,n]}[\xi_1\otimes\cdots\otimes\xi_n]=\rho\big(\bbE_{\hat\xi_n}\circ\cdots\circ\bbE_{\hat\xi_1}(\bbone)\big).
\end{equation}
\end{definition}

\begin{remark}[Universal unitality of identity insertion]\label{rem:unital}
The identity intervention is the unique intervention whose transfer map is unital for every process in the class. A general deterministic intervention $\hat{\cM}$ does \emph{not} give $\bbE_{\hat{\cM}}(\bbone)=\bbone$: a nontrivial channel changes future statistics. The normalization condition $\rho\circ\bbE_{\hat{\cM}}=\rho$ is instead the abstract form of ``instrument outcomes sum to one'', and its proof is precisely the causality of the comb (a deterministic operation at the latest slot cannot change earlier statistics).
\end{remark}

For a fixed process, $\bbE_{\hat{\cM}}(\bbone)=\bbone$ holds iff $\hat{\cM}$ leaves the conditional future statistics unchanged. The universal quantifier in Remark~\ref{rem:unital} is essential: degenerate fixed processes can admit other unital interventions.\footnote{For an i.i.d.\ process, every deterministic channel has unital transfer map. This does not affect uniqueness under the universal, all-processes quantifier.}

The generating formula \eqref{eq:generating} is read from right to left: start with the memory in its distinguished state $\bbone$, apply one update per intervention in time order, and evaluate the resulting functional at the boundary state $\rho$. Two features matter for what follows. First, there is one map $\bbE_{\hat\xi}$ per intervention \emph{element}, so the transfer structure resolves not only which instrument was applied but also which outcome occurred. Summing over outcomes recovers the channel-level update. Second, the family $\{\bbE_{\hat\xi}\}$ is not a group or even a semigroup of physical channels on a fixed system: it is an operator-valued measure on the space of interventions, and its composition law encodes the process's memory. For channel interventions, the maps $\bbE_{\hat{\cM}}$ describe how the future is steered by what is done now. This is the object that classical theory reaches only through conditioning, and which here exists without any quantum conditional state.

\subsection{Construction for finite Markov order}

We now show that finite Markov order forces the memory to be finite-dimensional, with an explicit bound:

\begin{proposition}[Finite-dimensional representation]\label{prop:triple}
Let $\{\cT_\Lambda\}$ be a stationary quantum time series with Markov order $\ell$ relative to the fixed full-rank IC causal break of Sec.~\ref{sec:prelim}. Then there exists a generating triple with $\dim\cB\le d^{4\ell}$, with the matrix order and complete positivity understood using the intervention cone $\mathcal O$ of Definition~\ref{def:triple}.
\end{proposition}

\noindent\textbf{Proof roadmap.} Stationarity and identity-insertion consistency enter first through Proposition~\ref{prop:extension}, supplying the stationary infinite past and hence the order unit and invariant boundary state. Finite Markov order relative to the fixed IC break enters through the block expansion \eqref{eq:blockexpansion}, which bounds the span of conditional future functionals by $d^{4\ell}$ (Lemma~\ref{lem:finitedim}). The definition of these functionals on all future windows makes the update map independent of the chosen past and yields the time-ordered generating formula (Lemmas~\ref{lem:transfer} and \ref{lem:boundary}). Finally, the factorization \eqref{eq:markovorder}, full rank of the IC POVM, and restriction of the order tests to the measure-and-prepare/identity operational cone enter the ratio estimate that proves the order-unit, positivity, and state properties (Lemma~\ref{lem:order}). No coherent-intervention extension is assumed.

The bound has a transparent meaning: after an $\ell$-step causal break, everything the past can influence about the future passes through the $\ell$-step memory block, whose operator space has dimension $d^{4\ell}$. The memory of the process is therefore no larger than the block that mediates it. The construction makes this precise by identifying the memory space with the span of all \emph{conditional future functionals}: the past matters only through the future statistics it still controls.

In the following, we present the key lemmas entering the proof in this subsection: finite dimensionality of the functional space (Lemma~\ref{lem:finitedim}) and the operational order structure (Lemma~\ref{lem:order}), with further details provided in Appendix~\ref{app:transfer}. Well-definedness of the update map requires equality of functionals on every future window, while the boundary-state step requires the stationary infinite past. These hypotheses and conclusions are stated explicitly in Lemmas~\ref{lem:transfer} and \ref{lem:boundary} and summarized below.

For a finite past intervention sequence $\xi_P$ on $P=[-m+1,0]$, define the \emph{conditional future functional}
\begin{equation}
\Phi_{\xi_P}(\xi_1,\dots,\xi_n):=\cT_{[-m+1,\,n]}\big[\xi_P\otimes\xi_1\otimes\cdots\otimes\xi_n\big],
\end{equation}
a single consistent object on all future windows, and set $\cQ:=\spn\{\Phi_{\xi_P}\}$. The first step of the proof bounds the dimension of this space:

\begin{lemma}[Finite dimensionality]\label{lem:finitedim}
$\dim\cQ\le d^{4\ell}$.
\end{lemma}
\begin{proof}
Split a window $[1,n]$ ($n\ge\ell$) into $M=[1,\ell]$ and $F=[\ell+1,n]$ and apply the block expansion \eqref{eq:blockexpansion} at the cut $P\mid M\mid F$:
\begin{widetext}
\begin{equation}
\cT[\xi_P\otimes\xi_1\cdots\xi_n]=\sum_{x}\underbrace{\tr\big[\Upsilon_{P}^{(x)}\,\xi_P^{T}\big]}_{=:c_x(\xi_P)}\cdot\underbrace{\tr\Big[\big(\Upsilon_{F}^{(x)}\otimes\Delta_{M}^{(x)}\big)\,\big(\xi_n\otimes\cdots\otimes\xi_1\big)^{T}\Big]}_{=:g_x(\xi_1,\dots,\xi_n)},
\end{equation}
\end{widetext}
so $\Phi_{\xi_P}=\sum_x c_x(\xi_P)\,g_x$ with at most $\dim W=d^{4\ell}$ terms. For an overcomplete IC instrument, we first pass to a basis subfamily of the dual set. Biorthogonality forces linear independence, so the count is at most $\dim W$. More explicitly, the independent subfamily is chosen from the instrument Choi operators spanning $W$, and $\{\Delta_M^{(x)}\}$ is then its dual basis. Testing both sides of \eqref{eq:blockexpansion} against these spanning instrument operators shows that this rewriting preserves the decomposition. The labels $x$ thereafter refer to the common independent subfamily.
\end{proof}

This argument has three features worth making explicit. First, the bound concerns functionals on \emph{all} future intervention tuples, coherent or not: each $g_x$ is defined by the displayed contraction for arbitrary $\xi_1,\dots,\xi_n$, so the dimension bound is not restricted to any causal-break subspace. Second, the instrument-relativity of the Markov condition enters only through the choice of the cut $M$ and its dual set. The block expansion \eqref{eq:blockexpansion} is an identity of operators, so it constrains the response to arbitrary histories $\xi_P$ and arbitrary future interventions at once. Third, the dual set $\{\Delta_M^{(x)}\}$, which is not individually positive, enters only this counting argument. The positivity arguments below (Lemma~\ref{lem:order}) are instead based on the factorization \eqref{eq:markovorder}, whose factors are, up to positive scalars, partial inner products of positive operators and hence positive.

Two further steps complete the construction. Appendix~\ref{app:transfer} gives their full statements and proofs, including the normalization identity. First, prepending one intervention, $(\bbE_{\hat\xi}\Phi)(\xi_1,\dots,\xi_n):=\Phi(\xi,\xi_1,\dots,\xi_n)$, defines a linear map $\bbE_{\hat\xi}:\cQ\to\cQ$ that depends only on the functional $\Phi$ and not on any past generating it or on any instrument choice. It satisfies $\bbE_{\hat\xi}\Phi_{\xi_P}=\Phi_{\xi_P\otimes\xi}$, is linear in $\xi$, and obeys the time-ordered composition law $\bbE_{\hat\xi'}\circ\bbE_{\hat\xi}\,\Phi_{\xi_P}=\Phi_{\xi_P\otimes\xi\otimes\xi'}$ (Lemma~\ref{lem:transfer}). Second, with the boundary state $\rho(\Phi):=\Phi(\varnothing)$ and the order unit $\bbone_{\cQ}:=\Phi^{\mathrm{stat}}_\varnothing$, the conditional future functional of the stationary infinite past, which exists by Proposition~\ref{prop:extension}, the generating formula \eqref{eq:generating} holds and $\rho\circ\bbE_{\cI}=\rho$ (Lemma~\ref{lem:boundary}). It remains to equip the memory space with the order structure that makes the triple physical:

\begin{lemma}[Order structure]\label{lem:order}
Define the matrix order on $\cQ$ by testing on the \emph{operational test set} $\mathfrak{T}_{\mathrm{op}}$ (positive combinations of intervention tuples whose slots are causal-break elements or identity insertions), with matrix levels entrywise. With respect to this order, $\bbone_{\cQ}$ is an order unit, $\bbE_{\hat\xi}$ is CP for $\xi\ge0$ in $\mathcal O$, and $\rho$ is a state. Whether this operational order coincides with the pointwise order on all coherent intervention tuples is open (Remark~\ref{rem:coherent}). The identity-intervention transfer map $\bbE_{\cI}$ is moreover a contraction in the evaluation norm
\begin{equation}\label{eq:evalnorm}
\|\Psi\|_{\mathrm{eval}}:=\sup\big\{|\Psi(\cM_1,\dots,\cM_n)|\;\big|\;n\in\mathbb{N},\ \hat{\cM}_j\ \mathrm{CPTP}\big\},
\end{equation}
which is a genuine norm on $\cQ$ since CPTP Choi operators span $\cV$.
\end{lemma}

\begin{proof}
Crucial insights about this lemma are as follows with detailed proofs given in Appendix~\ref{app:transfer}.
Positivity of $\Phi$ is tested on the \emph{operational test set} $\mathfrak{T}_{\mathrm{op}}$: positive combinations of intervention sequences whose slots are causal-break elements or identity insertions. The key estimate is the \emph{ratio lemma}: for a break element $O^{(x_M)}$ on the first $\ell$ slots and arbitrary $\xi_{\mathrm{rest}}$,
\begin{widetext}
\begin{equation}\label{eq:ratio}
\Phi_{\xi_P}\big(O^{(x_M)}\otimes\xi_{\mathrm{rest}}\big)=\beta_{x_M}(\xi_P)\,w_{x_M}(\xi_{\mathrm{rest}}),\qquad
\bbone_{\cQ}\big(O^{(x_M)}\otimes\xi_{\mathrm{rest}}\big)=p^{\mathrm{stat}}(x_M)\,w_{x_M}(\xi_{\mathrm{rest}}),
\end{equation}
\end{widetext}
where $w_{x_M}(\xi_{\mathrm{rest}}):=\tr[\Upsilon_F^{(x_M)}\xi_{\mathrm{rest}}^T]\ge0$ is independent of the past by \eqref{eq:markovorder}, $\beta_{x_M}(\xi_P)=\tr[\Upsilon_P^{(x_M)}\xi_P^T]\ge0$, and $p^{\mathrm{stat}}(x_M)\ge\big(\min_a\lambda_{\min}(\Pi^{(a)})\big)^{\ell}>0$ because the IC POVM is full rank. The history $\xi_P$ may be any CP intervention tuple. Neither trace preservation nor any other normalization is required. For each fixed history, $R(\xi_P)=\max_{x_M}\beta_{x_M}(\xi_P)/p^{\mathrm{stat}}(x_M)$ is finite, and $\xi_P\mapsto c\xi_P$ gives $R\mapsto cR$ for $c>0$. Once $\xi_P$ is fixed, this constant is uniform over future tests and does not grow with the past length. Appending identity slots leaves the corresponding functional and its domination constant unchanged. The order-unit property needs only this elementwise domination, $-R\bbone_{\cQ}\le\Phi_{\xi_P}\le R\bbone_{\cQ}$ on the test order, not a bound uniform over all histories. Extension to positive combinations is by linearity. Dropping identity slots by identity absorption reduces any interleaved test tuple to a consecutive break-only tuple, without changing either functional value or the domination constant. Identities at the beginning, in the interior, and at the end are all covered. If at least $\ell$ breaks remain, the ratio lemma applies to the first $\ell$. Otherwise the reduced tuple is one of the finitely many partial-break patterns, each with a finite, history-dependent ratio by the full-rank lower bound. Enlarging the constant if necessary therefore covers all of $\mathfrak{T}_{\mathrm{op}}$, independently of the original tuple length. Finally, because $\cI$ is a channel, prepending it maps every CPTP test tuple to another CPTP test tuple. The supremum defining \eqref{eq:evalnorm} therefore cannot increase under $\bbE_{\cI}$. This proves contractivity, the only norm property used by the spectral arguments below.
\end{proof}

\begin{remark}[Operational cone, matrix order, and boundedness]\label{rem:opcone}
Write $\cQ_+$ for the cone defined by $\mathfrak{T}_{\mathrm{op}}$. It is a proper cone: convexity and closedness are immediate from the defining tests, and pointedness follows because tensor products of causal-break and identity Choi operators span every finite-window intervention space. The ratio lemma gives two-sided domination of every self-adjoint element by a finite multiple of $\bbone_{\cQ}$. Thus $\cQ_+$ is generating, the order is Archimedean, and $\bbone_{\cQ}$ lies in the interior of $\cQ_+$. Equivalently, the order-unit norm is finite on every self-adjoint element. At matrix level, $[\Psi_{ij}]\ge0$ means $[\Psi_{ij}(\tau)]\ge0$ entrywise for every $\tau\in\mathfrak{T}_{\mathrm{op}}$. The test set therefore plays the dual role that defines the operational matrix order. It is not identified with the full pointwise order on coherent interventions. Correspondingly, complete positivity of $\bbE_{\hat\xi}$ is asserted exactly for elements satisfying $\xi\ge0$ in $\mathcal O$. Finite dimensionality makes all norms on $\cQ$ equivalent, and evaluation-norm contractivity makes $\bbE_{\cI}$ power-bounded, which is the boundedness input used throughout Sec.~\ref{sec:spectral}.
\end{remark}

\begin{remark}[Coherent interventions]\label{rem:coherent}
The operational cone is generated by measure-and-prepare interventions, which is also what process tensor tomography implements. For \emph{coherent} interventions entangled across the memory block, the ratio argument can fail: the memory-block factors $m_x=\tr[\Delta^{(x)}\xi_M^T]$ are signed, and destructive interference among memory sectors can make the stationary denominator anomalously small. Whether the order-unit domination extends to coherent test tuples (equivalently, whether the operational order coincides with the pointwise order on the full cone) is an open problem. See Conjecture~\ref{conj:coherent}. The generating formula, the evaluation norm, the order structure on the operational cone, and the spectral conclusions of Sec.~\ref{sec:spectral} do not depend on this extension.
\end{remark}

\subsection{Anchor cases}

Two degenerate cases calibrate the construction. For a classical order-$\ell$ Markov chain with transition matrix $P$, the memory space $\cQ$ is the function space on the last $\ell$ symbols, $\bbE_\cI$ is the shift-by-one-step map, $\bbE_{\hat{\cM}}$ for a measure-and-prepare instrument is the classical ``observe, then write'' update, $\rho$ is the stationary measure, and the ratio bound \eqref{eq:ratio} is the classical estimate $\mu/\pi\le1/\pi_{\min}$. Every theorem below reduces in this case to a standard fact about Markov chains. For a memoryless quantum process ($\ell=1$, IC preparations), $\cQ$ is dual to the system state space and $\bbE_{\hat{\cM}}$ is the (pre-)dual action of the channel $\hat{\cM}$: the transfer map is the channel itself, iterated. The genuinely new content of the construction lies between these anchors, where the memory is quantum and the instruments are arbitrary.

\section{Spectral theory: ergodicity and mixing}
\label{sec:spectral}

With the transfer map in hand, the classical spectral questions can be posed in earnest, and this section answers them: we determine when the process forgets, and at what rate. The answer comes in two layers, because quantum mechanics supplies two inequivalent readings of ``forgetting''.

\subsection{Two autocorrelations}

Classically there is one autocorrelation function. Quantumly there are two levels, governed by two different operators: the \emph{record-level} ACF of a fixed instrument (governed by $\bbE_{\cJ}:=\sum_i\bbE_{\hat{\cM}_i}$), and the following instrument-independent notion:

\begin{definition}[Idle-gap correlation]\label{def:idlegap}
For finite intervention blocks $\xi_A,\xi_B$, the \emph{idle-gap correlation} is
\begin{equation}\label{eq:idlegap}
C_{\xi_A,\xi_B}(\tau):=\rho\Big(\bbE_{\hat\xi_B}\circ\bbE_{\cI}^{\,\tau}\circ\bbE_{\hat\xi_A}(\bbone)\Big)-\rho\big(\bbE_{\hat\xi_B}(\bbone)\big)\,\rho\big(\bbE_{\hat\xi_A}(\bbone)\big).
\end{equation}
\end{definition}

The blocks $\xi_A,\xi_B$ may be arbitrary intervention elements. The informative case is \emph{outcome} blocks (trace-nonincreasing CP elements of an instrument). We fix the meaning of the centering by two calibrations. For an outcome block $\xi$ one has $\rho(\bbE_{\hat\xi}(\bbone))=\bbone_{\cQ}(\xi)$, the stationary occurrence probability of the block, so the subtracted term is the product of the stationary one-block marginals: $C_{\xi_A,\xi_B}(\tau)$ is the covariance of two occurrence events separated by an idle gap, and no separate centering convention is needed. For deterministic (trace-preserving) blocks $\rho(\bbE_{\hat\xi}(\bbone))=1$ and the correlation vanishes identically, as there is no randomness to correlate, so the definition is used with outcome or general sub-normalized blocks. Arbitrary elements are covered by linearity. Definition~\ref{def:ergodicity} below quantifies over all such blocks. The record-level ACF is the same construction with $\bbE_{\cI}$ replaced by the instrument-dressed map $\bbE_{\cJ}$, and its events are classical record events throughout.

The idle-gap correlation asks what the process remembers when nothing is done to it: two interventions are separated by $\tau$ slots of identity insertions, so the decay of $C_{\xi_A,\xi_B}(\tau)$ measures the intrinsic forgetting rate of the dynamics. The record-level ACF instead measures what survives when every slot is actively probed by the same instrument. The difference between $\bbE_{\cI}$ and $\bbE_{\cJ}$ is precisely the difference between ``not measuring'' and ``measuring and forgetting''. Their spectral difference is a quantitative expression of measurement backaction. Classically the two readings coincide for a pure readout (non-repreparing) instrument, because such observation leaves the state unchanged. They need not coincide for a measure-and-reprepare instrument: classical repreparation also changes the dynamics, as the key-locked counterexample of Appendix~\ref{app:keylock} demonstrates. For pure readouts, the splitting is the first place where the quantum spectral theory departs from the classical one. A related separation at the level of Markovianity, rather than mixing rates, was established by Taranto and Milz \cite{TarantoMilz2023}: records can be Markovian while the process fundamentally requires memory. With the two correlation levels in hand, the classical notions of ergodicity and mixing translate directly:

\begin{definition}[Ergodicity and exponential mixing]\label{def:ergodicity}
A stationary quantum time series is \emph{ergodic} if $\frac{1}{N}\sum_{\tau=0}^{N-1}C_{\xi_A,\xi_B}(\tau)\to0$ for all intervention blocks, and \emph{exponentially mixing} if $|C_{\xi_A,\xi_B}(\tau)|\le K_{\xi_A,\xi_B}\lambda^\tau$ for some $\lambda<1$.
\end{definition}

The quantification over arbitrary intervention elements is a linear extension, not a claim that every such element is separately realizable as a measurement. Physical event covariances are evaluated on instrument outcome blocks. Because CPTP Choi operators span $\cV$, the linear extension leaves the spectral conclusions unchanged.

As in the classical theory, ergodicity is the minimal demand (time-averaged correlations wash out), while mixing is the stronger demand that they decay at an exponential rate. The two are genuinely different quantumly as well as classically, and the difference will be visible in the spectrum.

\subsection{The ergodicity criterion}

Finite dimensionality plus contractivity (Lemma~\ref{lem:order}) make $\bbE_{\cI}$ a power-bounded operator on a finite-dimensional space, so all peripheral Jordan blocks are trivial (peripheral semisimplicity) and the mean ergodic projection $P:=\lim_N\frac1N\sum_{\tau<N}\bbE_{\cI}^{\tau}$ exists \cite{vonNeumann1932,Krengel1985}. Throughout this section, a \emph{simple} eigenvalue means an algebraically simple one. Under power-boundedness this is equivalent to a one-dimensional fixed-point space. The correlation quantifier is over arbitrary finite intervention elements. By linearity this is exactly the probe class used in the two-sided spanning argument below. Outcome blocks supply the event-covariance interpretation, but do not restrict the spectral quantifier. The first spectral criterion characterizes forgetting on average:

\begin{theorem}[Ergodicity criterion]\label{thm:ergodic}
For a stationary process with a finite-dimensional generating triple as in Proposition~\ref{prop:triple}, the following are equivalent: (i) the eigenvalue $1$ of $\bbE_{\cI}$ is algebraically simple (equivalently here, the fixed-point space is $\mathbb{C}\,\bbone_{\cQ}$). (ii) the process is ergodic. (iii) $P=\bbone_{\cQ}\otimes\rho$.
\end{theorem}

\noindent\textbf{Proof roadmap.} Finite dimensionality and evaluation-norm contractivity of $\bbE_{\cI}$ give power-boundedness, exclude nontrivial peripheral Jordan blocks, and make the mean ergodic projection $P$ well defined. Stationarity enters as $\rho\circ\bbE_{\cI}=\rho$, which identifies the invariant readout in the rank-one projection. The quantification over all intervention blocks in Definition~\ref{def:ergodicity} enters in both directions: right spanning converts a second fixed point into a nonvanishing Ces\`aro correlation, while the rank-one form of $P$ makes every such averaged correlation vanish (Appendix~\ref{app:spectral}).

Operationally, the criterion says that a process forgets on average exactly when its idle dynamics carries no conserved content: any nontrivial fixed point of $\bbE_{\cI}$ is a memory direction that never decays, and ergodicity is the absence of all such directions. The projection $P$ plays the role of the stationary expectation, mapping every memory element to its stationary shadow.

\subsection{A trivial peripheral spectrum implies exponential mixing}

Ergodicity concerns forgetting on average: only the time-averaged correlations are required to wash out. The operationally decisive notion is stronger, that the correlations themselves decay at a computable rate. The next theorem says that the intrinsic process-memory rate is read off the idle spectrum: once the eigenvalue $1$ is simple and no other eigenvalue lies on the unit circle, every idle-gap correlation decays exponentially, and the subleading eigenvalue modulus sets the intrinsic process-memory time. The precise statement is the following:

\begin{theorem}[Gap implies mixing]\label{thm:gap}
For a process with a finite-dimensional generating triple as in Proposition~\ref{prop:triple}, if the eigenvalue $1$ of $\bbE_{\cI}$ is simple and no other eigenvalue lies on the unit circle (we call the peripheral spectrum \emph{trivial} in this case), the process is exponentially mixing: for every $\lambda$ strictly larger than the subleading modulus $|\nu_2|:=\max\{|\nu|:\nu\in\spec\bbE_{\cI},\ \nu\ne1\}$, one has $|C_{\xi_A,\xi_B}(\tau)|\le K_{\xi_A,\xi_B}\lambda^\tau$ (with $\lambda=|\nu_2|$ itself when the subleading eigenvalues are semisimple. Jordan blocks contribute polynomial factors, absorbed by the strict inequality).
\end{theorem}

\noindent\textbf{Proof roadmap.} Simplicity of the eigenvalue $1$ and triviality of the remaining peripheral spectrum give the decomposition $\bbE_{\cI}=P+R$, with $P=\bbone_{\cQ}\otimes\rho$ supplied by Theorem~\ref{thm:ergodic}. Finite dimensionality converts the spectral radius of $R$ into the stated exponential operator bound, including the Jordan-block qualification. Substitution into Definition~\ref{def:idlegap} then proves decay for every pair of intervention blocks, with the block-dependent constant coming only from the readout and excitation factors (Appendix~\ref{app:spectral}).

Thus a gapped idle spectrum gives an \emph{intrinsic process-memory rate}: every idle-gap correlation is bounded using $|\nu_2|$, independently of the measurement scheme. The corresponding intrinsic process-memory time is a property of $\bbE_{\cI}$, not of an instrument-dressed record map.

\subsection{Exponential mixing implies a trivial peripheral spectrum}

The converse direction is what makes the spectral criterion a criterion rather than a sufficient condition: if every idle-gap correlation decays, then no peripheral mode can hide from all probes. Passing from correlation decay to the operator spectrum requires a two-sided spanning property of the construction, which we make precise after the statement:

\begin{theorem}[Mixing implies gap]\label{thm:converse}
For a process with a finite-dimensional generating triple as in Proposition~\ref{prop:triple}, if $C_{\xi_A,\xi_B}(\tau)\to0$ for all intervention blocks, then the peripheral spectrum of $\bbE_{\cI}$ is trivial (in the sense of Theorem~\ref{thm:gap}: the eigenvalue $1$ is simple and no other eigenvalue lies on the unit circle).
\end{theorem}

\noindent\textbf{Proof roadmap.} The hypothesis is decay for \emph{all} intervention blocks, not merely one selected pair. The right-spanning property makes these blocks excite every memory direction, and the left-spanning property makes their readouts detect every functional direction. Together they upgrade correlation decay to strong convergence $\bbE_{\cI}^{\tau}\to\bbone_{\cQ}\otimes\rho$. Power-boundedness then rules out nontrivial peripheral Jordan blocks, and any remaining nonunit peripheral eigenvector would produce a nondecaying oscillation detected by a suitable pair of blocks (Appendix~\ref{app:spectral}).

Together, Theorems \ref{thm:gap} and \ref{thm:converse} close the loop at the process level: triviality of the peripheral spectrum of the idle transfer map is \emph{equivalent} to exponential forgetting, so this spectrum determines the intrinsic long-time process memory and its intrinsic decay rate. Combined with Theorem~\ref{thm:ergodic}, which detects the weaker property of forgetting on average through the eigenvalue $1$ alone, this gives a complete spectral classification of the idle dynamics. One qualification matters in practice: a given intervention pair $\xi_A,\xi_B$ excites and reads only the modes it couples to, so an observed decay rate can be faster than the slowest mode. Sec.~\ref{sec:numerics} exhibits an instance where this distinction is essential.

\noindent\textbf{Two-sided spanning: from all correlations to the full operator.} Theorem~\ref{thm:converse} rests on a two-sided spanning property. This is the explicit bridge from decay of every observable correlation to a statement about the full operator spectrum, rather than a merely technical detail of the proof. The ambient space is the finite-dimensional $\cQ$ ($\dim\cQ\le d^{4\ell}$, Lemma~\ref{lem:finitedim}), equipped with the evaluation norm \eqref{eq:evalnorm}. Finite-dimensionality makes the norm choice immaterial. The construction satisfies two-sided spanning over \emph{all} finite intervention tuples: $\spn\{\bbE_{\hat\xi_1}\circ\cdots\circ\bbE_{\hat\xi_n}(\bbone_{\cQ})\}=\cQ$, because $\bbE_{\hat\xi_1}\circ\cdots\circ\bbE_{\hat\xi_n}(\bbone_{\cQ})=\Phi_{(\xi_1,\dots,\xi_n)}$ by Lemma~\ref{lem:transfer} and these functionals span $\cQ$ by definition. Similarly, $\spn\{\rho\circ\bbE_{\hat\xi_1}\circ\cdots\circ\bbE_{\hat\xi_n}\}=\cQ^*$, because these are the point evaluations at intervention tuples and CPTP tuples alone span $\cV$ (Lemma~\ref{lem:order}). Definition~\ref{def:ergodicity} quantifies over all intervention blocks, so correlation decay determines the strong operator limit of $\bbE_{\cI}^{\tau}$: the spectral criteria are intrinsic to the process, not relative to any restricted cone of probes. These three statements parallel Propositions 3.1 and 3.3 and Lemma 5.2 of \cite{FNW1992}. As in \cite[Example~4(2)]{FNW1992}, ergodicity does not imply mixing: a peripheral eigenvalue $-1$ produces nondecaying period-$2$ oscillations (see Sec.~\ref{sec:numerics} for a physical instance).

\subsection{The ergodic theorem for measurement records}

We now turn from what the process remembers to what an observer can learn. The practical question behind ergodic theory is whether a single experimental trajectory suffices for inference. The following theorem gives a sufficient condition with rates:

\begin{theorem}[Exponential mixing of records: LLN and CLT]\label{thm:record}
Fix a finite-outcome instrument $\cJ$ whose outcome elements lie in the operational cone (a measure-and-prepare instrument, for instance), and let $\bbE_{\cJ}:=\sum_i\bbE_{\hat{\cM}_i}$. Let $Y_t$ be the stationary record on this finite outcome alphabet, let $X_t=f(Y_t)$ for a real-valued function $f$ on the outcome alphabet, and set $S_N=\sum_{t=1}^N X_t$, $\mu=\mathbb{E}X_0$, $\gamma_k=\operatorname{Cov}(X_0,X_k)$, and $\sigma^2=\gamma_0+2\sum_{k\ge1}\gamma_k$. Suppose $\bbE_{\cJ}$ has a simple eigenvalue $1$ (with right fixed point $e_{\cJ}$, $\rho(e_{\cJ})=1$. Note that $e_{\cJ}\ne\bbone$ in general, since $\bbE_{\cJ}$ is not unital for a non-identity instrument) and no other peripheral eigenvalue. Then the following conclusions hold.
\begin{itemize}
\item[(i)] \emph{Record mixing.} The record is exponentially $\alpha$-mixing: for every $\lambda$ strictly larger than the subleading modulus $|\nu_2|:=\max\{|\nu|:\nu\in\spec\bbE_{\cJ},\ \nu\ne1\}$ of the dressed operator there is $K$ with
\begin{equation}
\alpha(n)\le K\lambda^n
\end{equation}
with $\lambda=|\nu_2|$ itself when the subleading eigenvalues are semisimple.
\item[(ii)] \emph{Law of large numbers.} $S_N/N\to\mu$ almost surely.
\item[(iii)] \emph{Central limit theorem.} The series defining $\sigma^2$ converges absolutely and $(S_N-N\mu)/\sqrt{N}\Rightarrow\mathcal{N}(0,\sigma^2)$ \cite{Rosenblatt1956,IbragimovLinnik1971,Bradley2005}. If $\sigma^2>0$, this is the usual nondegenerate CLT. If $\sigma^2=0$, the conclusion is degenerate: $(S_N-N\mu)/\sqrt{N}\to0$ in probability, so the centered fluctuations vanish at the $\sqrt{N}$ scale.
\end{itemize}
\end{theorem}

\noindent\textbf{Proof roadmap.} Process stationarity and comb normalization make the finite-alphabet record stationary and consistent. For record events, the indicator-weighted outcome maps $\Gamma_A$ satisfy $0\le\Gamma_A\Psi\le\bbE_{\cJ}^{\,m}\Psi$ for positive $\Psi$. Base-norm contractivity and the normalization identity $\rho\circ\bbE_{\cJ}=\rho$ make the event bounds uniform over events and interval lengths. The decomposition $\bbE_{\cJ}^{\,n}=e_{\cJ}\otimes\rho+R^n$ then separates the stationary component from the exponentially decaying remainder. Combining it with stationarity gives $\rho(\Gamma_B e_{\cJ})=\rho(\Gamma_B\bbone)=P(B)$, which identifies the limiting record marginal. These steps yield $\alpha(n)\le K\lambda^n$ and the covariance bound $|\gamma_k|\le C'\lambda^k$. Boundedness of $f$, stationarity, and the resulting summability bounds provide exactly the hypotheses of the classical $\alpha$-mixing LLN and CLT, including the degenerate case $\sigma^2=0$.

Physically, the theorem extends the classical single-trajectory ergodic principle to the classical record produced by a fixed quantum instrument. It does not replace the quantum process itself by a classical process. The subleading eigenvalue of $\bbE_{\cJ}$ controls an \emph{instrument-specific record rate} and the associated confidence intervals, distinct from the intrinsic process-memory rate governed by $\bbE_{\cI}$. The detailed proof is given in Appendix~\ref{app:spectral}.

As a direct consequence, single-trajectory estimation is available for the stated record-statistic class:

\begin{corollary}[Single-trajectory record inference]
On the class of stationary processes with finite Markov order relative to the fixed IC causal break, for any finite-outcome operational-cone instrument with a gapped record operator and any bounded real-valued function of its record, the sample mean $S_N/N$ is a strongly consistent estimator of $\mu$ from a single measurement trajectory, with the CLT of Theorem~\ref{thm:record} describing its asymptotic fluctuations. The record mixing rate is instrument-specific and controlled by the spectrum of $\bbE_{\cJ}$.
\end{corollary}

\begin{remark}[Relation to earlier limit theorems]
Theorem~\ref{thm:record} complements the Sanov/CLT results for quantum Markov chain outputs \cite{vanHorssen2015} and the limit theorems for quantum trajectories \cite{KM2004,Benoist2023,Benoist2025}, which correspond to memoryless environments. Here the memory is an arbitrary finite-order process tensor. Conversely, the trajectory literature covers conclusion types we do not address (large deviations, Berry--Esseen rates \cite{Benoist2025}) and does not require finite Markov order. The two settings cover different regimes.
\end{remark}

The preceding theorem describes when a record is statistically informative across long times. There is also a structural restriction on which perfectly preserved memory can enter the record at all. We state it at the channel level, in the Schr\"odinger picture, for the driven memory channel generated by the fixed instrument.

\begin{theorem}[Record-invisibility theorem for correctable noncommutative blocks]\label{thm:nbl}
Let $\Phi$ be the finite-dimensional driven memory channel generated by a fixed finite-outcome measure-and-reprepare instrument, and suppose that $\Phi$ has a faithful stationary state on the support of its peripheral algebra. Let $\mathcal A$ be a noncommutative simple direct summand of that peripheral algebra which is preserved as a block by $\Phi$, so that $\Phi|_{\mathcal A}=\mathrm{Ad}_V$ for a unitary $V$. Let $\mathcal C$ be the fixed channel from the memory to the measured system supplied by the readout collision (a fixed complementary marginal of $\Phi$), and let $\{F_i\}$ be the instrument's readout POVM. Then there are constants $p_i\ge0$, independent of the memory content in $\mathcal A$, such that
\begin{equation}\label{eq:nbl-readout}
\begin{aligned}
r_i(X)&:=\tr\!\big[F_i\,\mathcal C(X)\big]=\tr(X)\,p_i,\\
&\quad\text{for every }X\in\mathcal A\text{ and every outcome }i,
\end{aligned}
\end{equation}
with $\sum_i p_i=1$. Consequently, no finite-string record probability or bounded real-valued function of the record generated by this fixed readout can distinguish the content of $\mathcal A$: this perfectly preserved memory mode is record-invisible.
\end{theorem}

\noindent\textbf{Proof roadmap.} The detailed proof is given in Appendix~\ref{app:nbl}. The argument has three steps. First, the peripheral eigenstructure theorem gives the blockwise unitary action. Second, Petz recovery propagates reversibility through the channel composition. Third, complementarity forces the complementary channel to factor through the trace on $\mathcal A$. This trace factorization then yields the finite-string conclusion.

The complementarity theorem of Ref.~\cite{KKS2008} gives privacy to the environment. Here the environment is the measurement record, and the trace-factorization is the record-level form of that privacy. On each invariant simple block, the maximally mixed state is invariant under $\mathrm{Ad}_V$ and hence is faithful and stationary on that block. Accordingly, no extra support condition is needed for a channel with a full-rank stationary state on its whole memory space, or when the peripheral algebra consists only of this block. With several peripheral summands and no such full-rank state, positive stationary weights on every summand must be assumed explicitly, since finite Markov order alone does not supply them.

This is the quantum-specific content of the two-level separation. Exact unitary preservation of a noncommutative simple block and visibility to the fixed record cannot both hold. The statement does not forbid a destructive measurement of quantum memory. It forbids reading out its noncommutative content while the driven dynamics preserves that content perfectly. It is a time-series counterpart of the impossibility of broadcasting noncommuting states \cite{Barnum1996}, and has no analogue for classical memory, whose content can in principle be copied and read out.

\begin{remark}[Why the simple-block hypothesis is essential]\label{rem:nbl-scope}
Simplicity cannot be relaxed to mere noncommutativity. With several simple summands, the peripheral algebra has a nontrivial classical center, and a complementary channel may depend on the summand label even when each block evolves unitarily. Likewise, commutative peripheral algebras---including conserved classical bits and cyclic permutations---are not covered by Theorem~\ref{thm:nbl}. The key-locked memory of Appendix~\ref{app:keylock} is exactly such a classical visible mode. What is excluded is record dependence on the noncommutative content within a single perfectly preserved simple block.
\end{remark}

\section{Physical scenarios}
\label{sec:scenarios}

The objects above are abstract, but each has a direct physical reading. We collect three scenarios: physical realizations of the model class, a direct application of Theorem~\ref{thm:record}, and an illustrative control mechanism for the two-level independence.

\noindent\textbf{Collision models as the canonical platform.} A collision model, in which the system interacts sequentially with fresh ancillas, is both a standard model of open-system dynamics \cite{Ciccarello2022} and a standard physical realization of our finite-Markov-order class: the ancillas carry the memory, in the instance of Sec.~\ref{sec:numerics} the Markov order is the number of consecutive collisions each ancilla participates in, and the idle transfer map is the memory channel of the collision chain. Cavity QED with crossing atoms and superconducting qubits coupled to resettable ancilla resonators implement such models directly. Section~\ref{sec:numerics} uses a two-collision model. The ancilla stream is the environment, and its partial retention is the non-Markovianity.

\noindent\textbf{Continuous monitoring and single-run inference.} Continuously monitoring an open quantum system (homodyne or photodetection records of a driven qubit, for example) produces a measurement record of the kind Theorem~\ref{thm:record} governs. The theorem's content for this setting is practical: when the instrument-dressed operator has a gap, a single experimental run suffices to estimate bounded observables of that record, with the long-run variance and instrument-specific inference rate read off the record's autocorrelation and the spectrum of $\bbE_{\cJ}$. The idle-gap correlation of Definition~\ref{def:idlegap} is the process's memory kernel measured without touching it: it answers the question of what the dynamics remembers when left alone, which is exactly the quantity that environment characterization tries to access.

\noindent\textbf{Memory-assisted control and dissipation dodging.} The two-level asymmetry has a control reading. An instrument that counter-rotates against the process's internal rotation can keep a memory direction alive indefinitely (Sec.~\ref{sec:discussion}), so a dissipative process can nonetheless carry permanent record correlations. This is a dynamical-decoupling-type phenomenon seen from the process side: the process's own dissipation sets the baseline, and the instrument's coherence against it determines what survives. The example illustrates the independence of the two mixing levels. Its coherent-intervention regime is outside the operational-cone hypotheses of Theorem~\ref{thm:record}.

\section{Numerical illustration: a collision model with Markov order 2}
\label{sec:numerics}

We now illustrate the spectral criteria of Sec.~\ref{sec:spectral}, in particular the mixing criterion of Theorem~\ref{thm:gap} and the record ergodic theorem, Theorem~\ref{thm:record}, on an exactly solvable model. We consider a qubit collision model in which each ancilla, fresh in $|0\rangle$, collides with the system at two consecutive steps before being discarded, so that the Markov order is $2$ and the memory is carried by a single ancilla qubit. We test two collision gates, $U(\theta)=e^{-i\theta X\otimes X}$ and the extended gate $U_{\mathrm{ext}}(\theta)=e^{-i\theta(XX+YY/2)}$. The computations combine analytic chain formulas with dense double-precision diagonalization and have no Monte Carlo element. Trace preservation, probability normalization, validity of the stationary memory state, and the i.i.d.\ limit $\theta\to0$ were checked to machine precision (Appendix~\ref{app:numerics}). These exact formula evaluations and numerical diagonalizations are consistency checks of the implementation of the stated formulas, not independent validations of the general mathematical theorems.

\noindent\textbf{Record level (Theorem~\ref{thm:record}).} With the instrument ``measure $\sigma_z$, reprepare $|+\rangle$'', the record operator $M$ on the ancilla memory was computed as a $4\times4$ superoperator. For the base gate, $M$ is provably rank one for \emph{any} measure-and-prepare instrument ($\nu_2=0$. The record is at most one-step correlated): the ancilla's written components and read-out direction are misaligned for single-Pauli gates. For the extended gate, the record ACF decays geometrically over $4$--$10$ orders of magnitude, and the fitted instrument-specific record rate agrees with $|\nu_2(M)|$ to five to six decimal places:

\begin{center}
\begin{tabular}{c|c|c}
$\theta$ & $|\nu_2(M)|$ & fitted ACF rate\\\hline
$0.2$ & $0.005985$ & $0.005985$\\
$0.4$ & $0.078037$ & $0.078038$\\
$0.6$ & $0.276959$ & $0.276959$\\
$0.8$ & $0.514161$ & $0.514161$
\end{tabular}
\end{center}

\begin{figure*}[t]
\centering
\includegraphics[width=\textwidth]{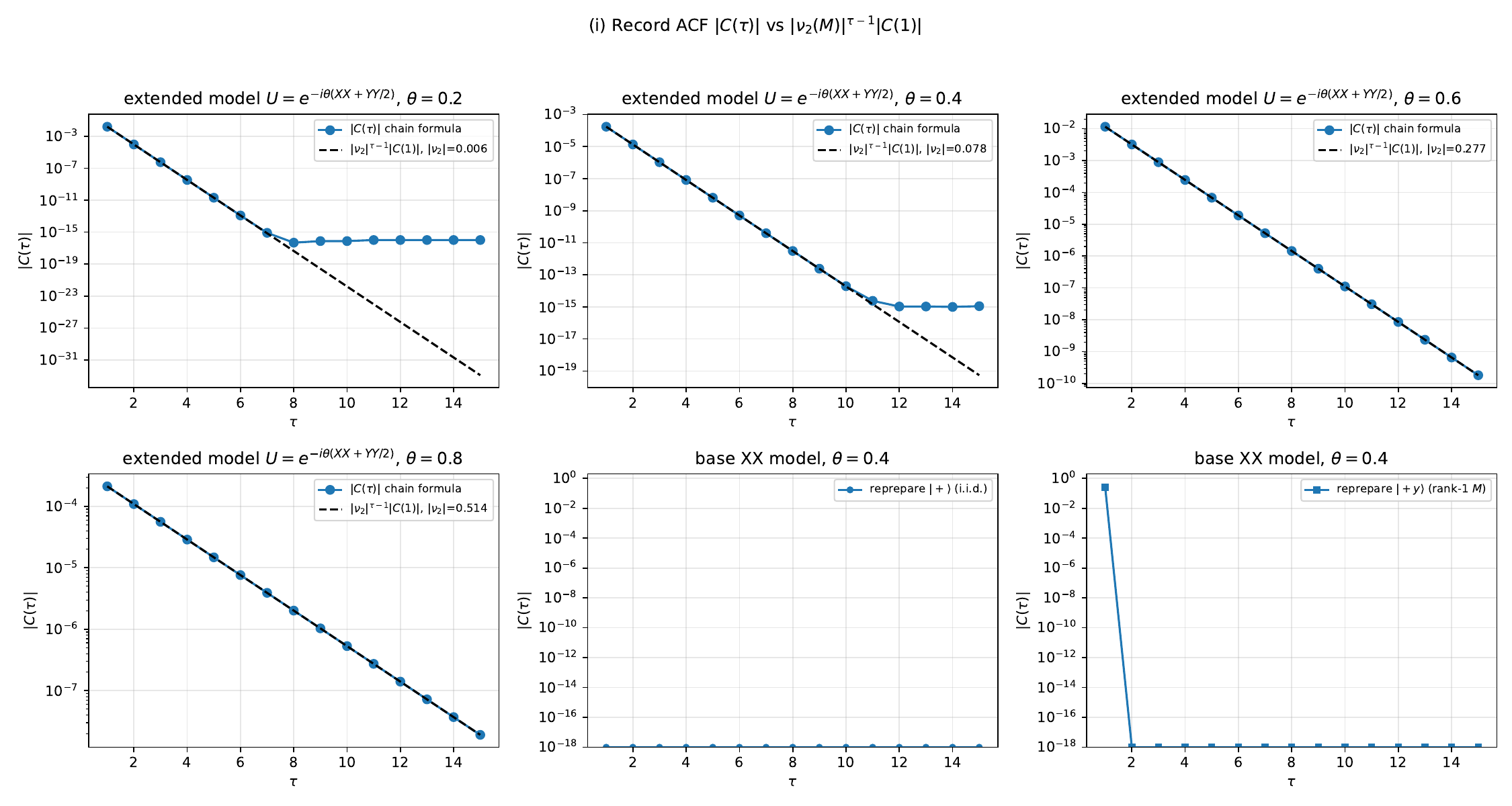}
\caption{Consistency check of the model-specific closed form of the record autocorrelation for the extended gate $U_{\mathrm{ext}}(\theta)=e^{-i\theta(XX+YY/2)}$ at $\theta=0.2,0.4,0.6,0.8$: the record autocorrelation $|C(\tau)|$ (markers) versus $|\nu_2(M)|^{\tau-1}|C(1)|$ (dashed curves), where $M$ is the $4\times4$ record operator on the one-ancilla memory and $|\nu_2(M)|$ its subleading eigenvalue modulus. The bottom-middle and bottom-right panels show the base gate $U(\theta)=e^{-i\theta X\otimes X}$ with repreparation in $|+\rangle$ and $|+y\rangle$, respectively. Its record operator is rank one. The plateau at $\sim10^{-15}$ is the floating-point noise floor. Analytic chain formulas together with double-precision diagonalization give a model-specific consistency check, not an independent test of the general mixing bound in Theorem~\ref{thm:record}.}
\label{fig:record}
\end{figure*}

\noindent\textbf{Idle level (Theorem~\ref{thm:gap}).} The idle step map $N$ on the joint system--ancilla memory has the closed-form nonzero spectrum $\{1,1,\cos4\theta,\cos4\theta\}$ (numerically matched to $\le4\times10^{-15}$), and the idle-gap correlation of a $z$-preparation/$z$-measurement pair is exactly $\Delta(\tau)=-\cos^2(2\theta)\,(\cos4\theta)^{\tau-1}$. The twofold degenerate eigenvalue $1$ (a conserved $\sigma_x$ functional) means the idle process is \emph{not} mixing. The observed intrinsic process-memory decay rate for this probe pair is that of the fastest nontrivial mode \emph{coupled to} the chosen preparation/measurement pair, an instance in which the hypotheses of Theorem~\ref{thm:gap} fail and the conclusion must be read mode by mode. At $\theta=\pi/4$ the peripheral eigenvalue $-1$ appears (period-$2$ oscillation), a physical analogue of \cite[Example~4(2)]{FNW1992}.

\begin{figure*}[t]
\centering
\includegraphics[width=\textwidth]{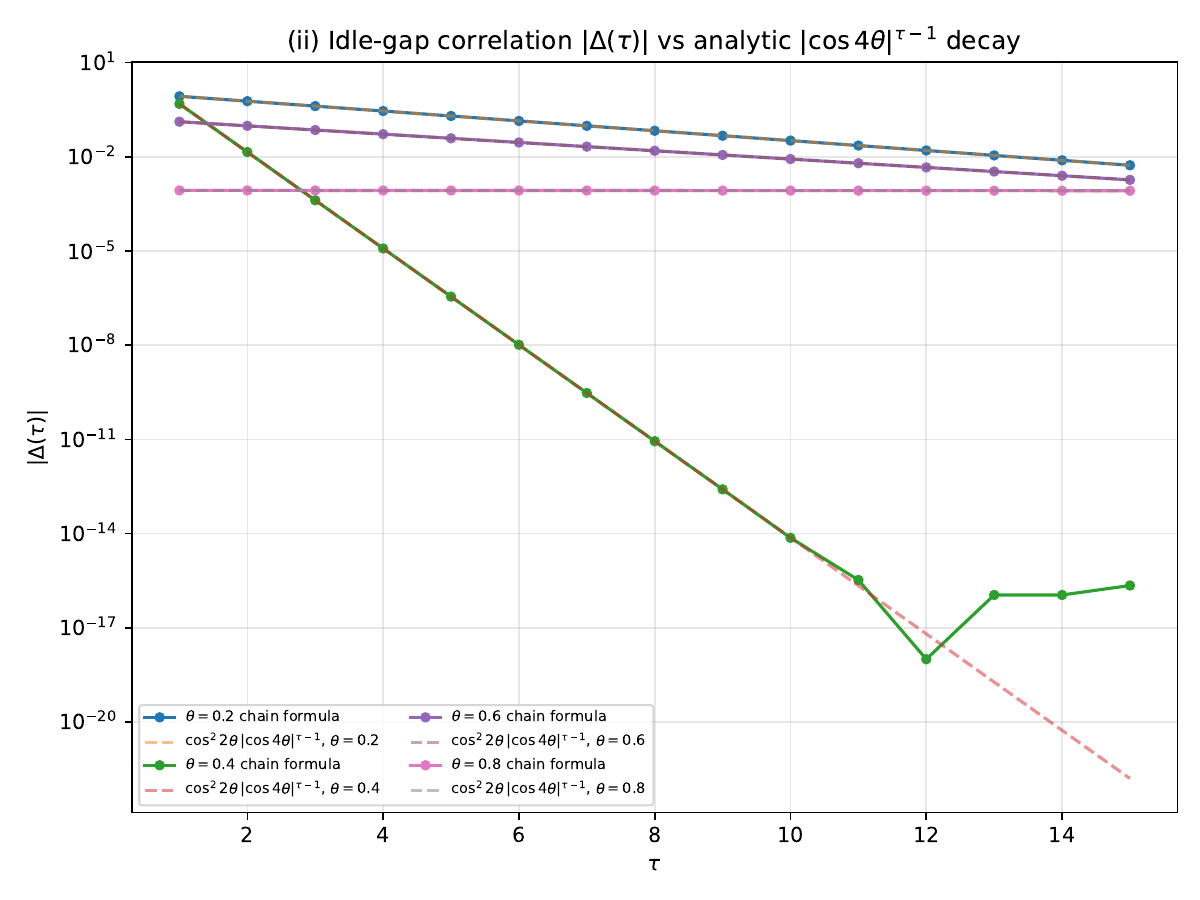}
\caption{Mode-by-mode consistency check of the model-specific closed form of the idle-gap correlation for the base gate at $\theta=0.2,0.4,0.6,0.8$: $\Delta(\tau)$ for a $z$-preparation/$z$-measurement pair (markers) versus $-\cos^2(2\theta)(\cos4\theta)^{\tau-1}$ (curves). In this example the hypotheses of Theorem~\ref{thm:gap} do not hold: the eigenvalue $1$ is twofold degenerate and the idle process is non-mixing. The figure therefore checks the closed form mode by mode rather than testing that theorem. The sign-alternating cases $\theta>\pi/8$ are visible. $\theta=0.8$ is close to the period-$2$ point $\theta=\pi/4$, where the eigenvalue $-1$ enters the peripheral spectrum of $N$. Below $\sim10^{-16}$ the signal is at the double-precision noise floor. Analytic chain formulas are compared with double-precision diagonalization.}
\label{fig:idle}
\end{figure*}

\begin{figure*}[t]
\centering
\includegraphics[width=\textwidth]{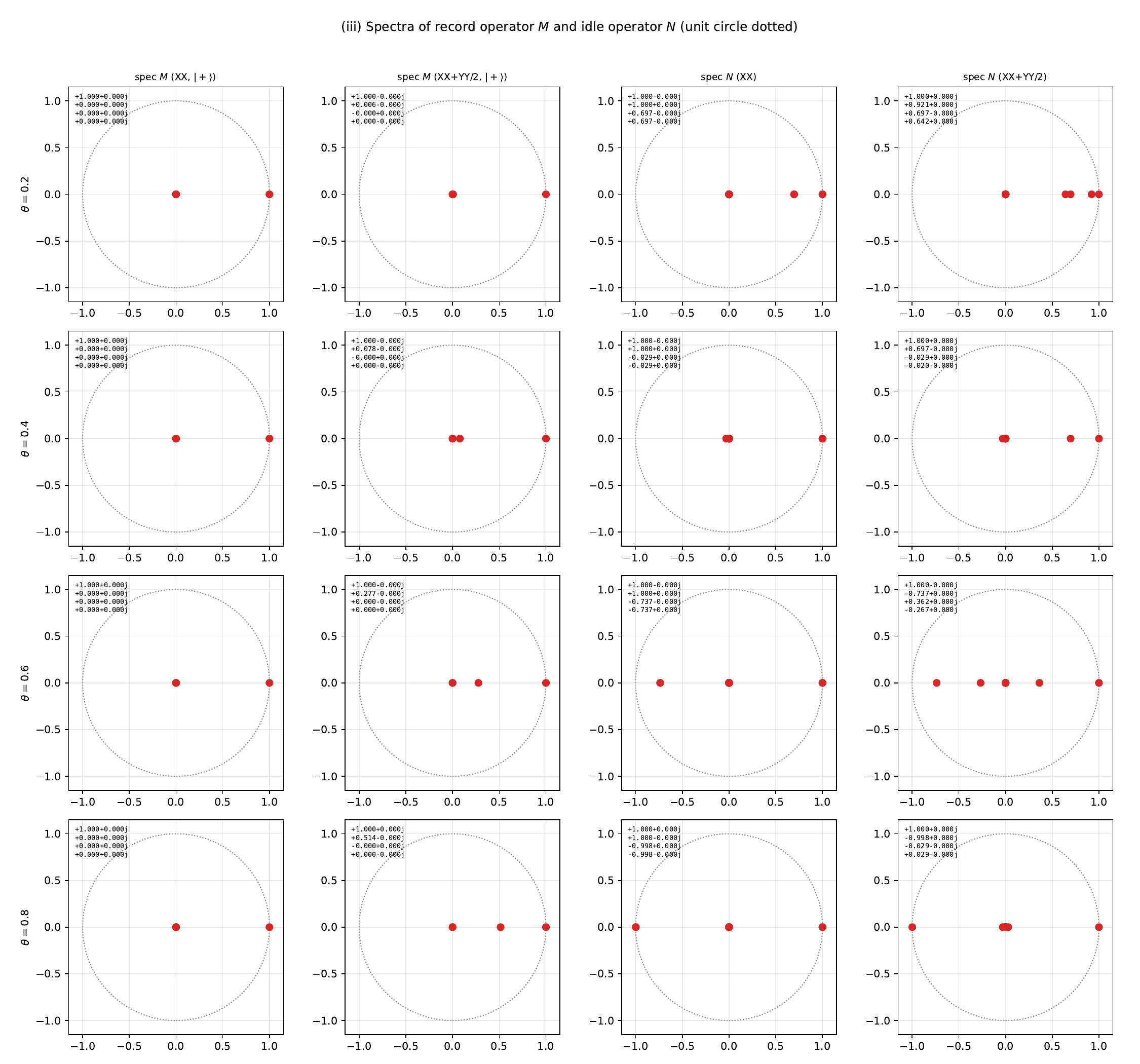}
\caption{The spectra behind both levels: eigenvalues of the idle step map $N$ and of the record operator $M$ in the complex plane, for the two gates and the tested couplings (see text for details). The degenerate eigenvalue $1$ of $N$ (the conserved $\sigma_x$ functional) and the peripheral approach at $\theta=\pi/4$ are visible. The record operator's spectrum is strictly inside the unit disk for the extended gate and degenerates to $\{1,0,0,0\}$ for the base gate.}
\label{fig:spectra}
\end{figure*}

\noindent\textbf{Cross-level comparison.} At all tested points, $\nu_2(M)\le\nu_2(N)$. Equivalently, the tested measure-and-prepare instruments increase contractivity at those points. This is an observation at all tested points, not a general theorem. Idle mixing implied record mixing at every tested point (including the nontrivial extended gate), while the converse is false for the base gate: the record is i.i.d.\ while the idle process fails to mix, because the instrument destroys the conserved $\sigma_x$ quantity. The point spectra are not nested. Any general relation must concern the peripheral spectrum or mixing, not the full spectrum.

\noindent\textbf{Reproducibility.} All quantities above are finite-dimensional and computed from analytic chain formulas together with double-precision diagonalization. With the gate, the reprepare state, and the instrument fixed as stated, the record operator $M$ ($4\times4$, acting on the one-ancilla memory) and the idle step map $N$ ($16\times16$, acting on the joint system--ancilla memory) are given by the explicit formulas of Appendix~\ref{app:numerics}. The stationary memory state $\eta_*$ is the fixed point of $M$, with Bloch vector $(0,-\sin2\theta,\cos2\theta)$. Record probabilities and autocorrelations follow from the chain rule stated there. There is no Monte Carlo element: the only numerical steps are dense matrix algebra and diagonalization in double precision, with the sanity checks of Appendix~\ref{app:numerics} passed to $\sim10^{-16}$. Fitted rates are extracted from the log-linear decay of $|C(\tau)|$ over the window between the onset of asymptotic decay and the floating-point floor near $10^{-15}$ (visible in Fig.~\ref{fig:record}). The agreement quoted in the table is to five to six decimal places. At and near the period-$2$ point $\theta=\pi/4$, where the peripheral eigenvalue $-1$ makes any finite-window fit meaningless, the spectra are read from the closed forms of Appendix~\ref{app:numerics} and the numerical diagonalization is a cross-check.

Taken together, these computations check the model-specific analytic chain formulas and their double-precision diagonalization at both correlation levels, with agreement to five to six decimal places at the tested points. The model exhibits the two-level asymmetry and illustrates that an observed decay rate reflects the modes coupled to the chosen probes: an intrinsic process-memory rate is read from $N$ (the realization of $\bbE_{\cI}$), whereas an instrument-specific record rate is read from $M$ (the realization of $\bbE_{\cJ}$).

\section{Discussion and open problems}
\label{sec:discussion}

We have built the stationary theory of quantum time series on the process-tensor foundation and proved structural theorems for the finite-order model class: the definition (Definition~\ref{def:stationary}) with its existence theorem (Proposition~\ref{prop:extension}), the finite-dimensional causal transfer map (Proposition~\ref{prop:triple}), the two-level autocorrelation structure (Definition~\ref{def:idlegap}), and the spectral criteria with the record ergodic theorem (Theorems \ref{thm:ergodic}--\ref{thm:record}). However, our results are established under two restrictions: finite Markov order relative to an IC causal break, and measure-and-prepare operational cones. Whether the spectral criteria persist when either restriction is relaxed, for infinite-order memory on the one hand and for coherent interventions on the other, is open. Beyond these, several further questions are already visible, and we here list five candidates:

\begin{enumerate}
\item[(i)] \emph{Coherent interventions and temporal interference.} The order-unit property was established on the operational cone (measure-and-prepare plus identity). Whether it extends to the full cone of coherent interventions is open:
\begin{conjecture}\label{conj:coherent}
There is a dichotomy between processes for which the operational order extends to coherent interventions and processes for which it fails. We conjecture that the failure class is nonempty and exhibits an interference phenomenon among memory sectors with no classical counterpart.
\end{conjecture}
We do not currently have a decidable criterion for the failure class. Finding one (for instance, a spectral or entropic quantity of the transfer map that detects it) is itself part of the problem. This boundary has no classical counterpart: it marks where temporal interference (interference among memory sectors) becomes visible in asymptotic statistics.
\item[(ii)] \emph{Cross-instrument spectra: the two levels are independent.} The independence of the two levels is a general structural feature of the framework, shared by classical and quantum models. Record mixing does not imply process mixing (Sec.~\ref{sec:numerics}: instruments can destroy conserved quantities of the idle dynamics). Process mixing does not imply record mixing either. Consider the single-qubit Markov channel $\Lambda=R_y(\pi/2)\circ C$, where $C(\rho)=\frac{1+\lambda}{2}\rho+\frac{1-\lambda}{2}X\rho X$ ($0<\lambda<1$): the Bloch matrix of $\Lambda$ has eigenvalues $\{\lambda,\pm i\sqrt{\lambda}\}$, so $\Lambda$ is mixing, but composing with the unitary instrument $\mathcal{N}=R_y(-\pi/2)$ gives $\Lambda\mathcal{N}$ two fixed states $|0\rangle\langle0|$ and $|1\rangle\langle1|$. With a weak $\sigma_z$ readout of strength $\varepsilon$, the record autocorrelation is exactly constant, $C(\tau)=4\varepsilon^2$ for all $\tau\ge1$, while the idle process mixes at the intrinsic process-memory rate $\sqrt{\lambda}$ (a direct Bloch-sphere computation). The instrument rotates the memory content away from the contracting direction at every step, dodging the dissipation that drives idle mixing. This coherent-instrument example illustrates the two-level independence. It lies outside the operational-cone hypotheses of Theorem~\ref{thm:record} and is not covered by that theorem. The forward implication fails even for causal breaks: a classical permutation gate uses the reprepared state as a key that keeps a classical memory channel open (record non-mixing) while the idle evolution forgets in two steps (process mixing), as shown self-containedly in Appendix~\ref{app:keylock}. The illustrative phenomenon which cannot happen on the classical side is precisely the no-broadcasting phenomenon of Theorem~\ref{thm:nbl}: perfectly preserved noncommutative simple-block memory is invisible to the record generated by a fixed measure-and-reprepare instrument. The proof in Appendix~\ref{app:nbl} combines peripheral eigenstructure \cite{WolfPG2010}, Petz recovery through the collision composition \cite{Petz1986,Petz1988,Jencova2024}, and complementarity of the readout channel \cite{KKS2008}. Classical central content and commutative peripheral modes remain visible, as the key-locked example demonstrates. Classical memory content can in principle be copied and read out \cite{Barnum1996}. Theorem~\ref{thm:nbl} covers the perfectly preserved case. The classification for partially preserved blocks is open. Already at the channel level, mixing channels are not closed under right composition.
\item[(iii)] \emph{Peripheral structure.} The full analogue of \cite[Prop.~3.3]{FNW1992} (cyclic peripheral spectrum and periodic decomposition) requires Perron--Frobenius theory on operator systems \cite{EvansHoeghKrohn1978,Wolf2012} and is left to future work.
\item[(iv)] \emph{Uniqueness.} Whether two-sided minimal generating triples are unique up to linear isomorphism remains open, as does its finitely-correlated-state counterpart for general $C^*$-FCS \cite{FNW1992}.
\item[(v)] \emph{Outlook.} The representation and spectral theory built here are the scaffolding for a quantum time series analysis: spectral (Wiener--Khinchin-type) theory, ARMA-type parametrized model families, estimation and order selection, and nonclassicality measures of temporal statistics are the natural next steps, and the criteria proved here are what those developments will stand on.
\end{enumerate}

\noindent\textbf{Data availability.} The data supporting the numerical results are contained within the article. The analytic formulas used to generate these results are given in Appendix~\ref{app:numerics}.

\begin{acknowledgments}
This work is supported by the National Natural Science Foundation of China [grant number 12401158] and the Fundamental Research Funds for the Central Universities [grant number 30925010422].
\end{acknowledgments}

\newpage 
\onecolumngrid
\input{appendix}

\end{document}

%% file: appendix.tex
\appendix

\section{Details for Sec.~\ref{sec:stationary}: the covariant extension}
\label{app:stationary}

This appendix expands the proof of Proposition~\ref{prop:extension} (unique covariant extension of a stationary comb family to the two-sided infinite time axis), whose proof sketch is given in Sec.~\ref{sec:stationary}. We use the finite-dimensional, lift-based GET of Milz et al.\ \cite{Milz2020}. The hypotheses used here are: (i) every local intervention space is finite-dimensional. (ii) For each finite $\Lambda$, $\cK_\Lambda$ is the causally compatible intervention domain and $\cK^\sharp_\Lambda$ is its lifted domain, with identity insertions outside the finite support. (iii) Whenever $\Lambda\subseteq\Lambda'$, the combs agree under restriction to the lifted compatible domain, equivalently under identity insertion at the slots of $\Lambda'\setminus\Lambda$. Definition~\ref{def:stationary}(i) supplies exactly this consistency, for intervals and finite unions of intervals and for dropped slots in the interior as well as at the boundary. The finite-dimensional and causal-compatibility requirements are already part of the comb setting. The GET therefore associates to $\{\cT_{\Lambda_k}\}$ a unique comb $\overline{\cT}^\sharp$ on the closure $\overline{\cK}^\sharp_\Lambda$ of the lifted finitely-supported interventions. For $\Lambda=\mathbb{Z}$ with its shift $\sigma$, define $\cT'[\xi]:=\overline{\cT}^\sharp[\sigma(\xi)]$. The proof is divided into three steps.

\textbf{Step 1: $\sigma$ preserves the domain.} An element of $\cK^\sharp_{\mathbb{Z}}$ has the form $\xi=\xi_{\Lambda_k}\otimes\bigotimes_{t\notin\Lambda_k}\cI_t$ with $\xi_{\Lambda_k}\in\cK_{\Lambda_k}$. A supported window $[a,b]$ is mapped to $[a+1,b+1]$, and a finite union of intervals is mapped in the same way. Thus the shifted tuple is $\sigma(\xi_{\Lambda_k})\otimes\bigotimes_{t\notin\Lambda_k+1}\cI_t$. Since the whole family is defined with respect to one fixed temporal causal order, and $\cK_{\Lambda_k}$ is defined by compatibility with that order, $\sigma(\cK_{\Lambda_k})=\cK_{\Lambda_k+1}$. Hence finite-interval lifting and causal compatibility are both preserved, $\sigma(\cK^\sharp_{\mathbb{Z}})=\cK^\sharp_{\mathbb{Z}}$, and by continuity $\sigma(\overline{\cK}^\sharp_{\mathbb{Z}})=\overline{\cK}^\sharp_{\mathbb{Z}}$.

\textbf{Step 2: finite marginals agree.} For any finite interval $[a,b]$, the shifted lifted tuple above has finite marginal $\cT'^{|[a,b]}=\cT_{[a+1,b+1]}\circ\sigma=\cT_{[a,b]}$ by Definition~\ref{def:stationary}(ii). The same identity holds on finite unions of intervals by lifting and identity-insertion consistency.

\textbf{Step 3: uniqueness.} The GET states that a comb on $\overline{\cK}^\sharp_{\mathbb{Z}}$ is uniquely determined by its finite marginals. Applying this to $\cT'$ and $\overline{\cT}^\sharp$ this gives equality. \qed

\begin{remark}
The GET's uniqueness statement concerns the functional on $\overline{\cK}^\sharp_{\mathbb{Z}}$. The underlying process on the larger container space need not be unique, but the ambiguity is experimentally undetectable \cite{Milz2020}. All representations in this paper are representations of $\overline{\cT}^\sharp$.
\end{remark}

\section{Details for Sec.~\ref{sec:transfer}: the causal transfer map}
\label{app:transfer}

This appendix completes the construction behind Proposition~\ref{prop:triple}: it verifies the consistency of the conditional future functionals, proves Lemma~\ref{lem:finitedim} in detail, states and proves Lemmas~\ref{lem:transfer} and \ref{lem:boundary} (which the main text summarizes in prose), gives the full proof of Lemma~\ref{lem:order}, and works out the anchor cases of the generating triple.

\subsection{Consistency of the conditional future functionals}

This subsection verifies the consistency property invoked in Sec.~\ref{sec:transfer} when each $\Phi_{\xi_P}$ is treated as a single object on all window lengths. That property follows from consistency (Definition~\ref{def:stationary}(i)): for $n<n'$, $\Phi_{\xi_P}(\xi_1,\dots,\xi_n)=\Phi_{\xi_P}(\xi_1,\dots,\xi_n,\cI,\dots,\cI)$, since inserting identity interventions marginalizes. Linearity of $\Phi_{\xi_P}$ in each argument follows from multilinearity of $\cT$.

\subsection{Proof of Lemma~\ref{lem:finitedim} (finite dimensionality)}

Fix the IC causal break $\{O^{(x_M)}\}$ and its dual set $\{\Delta_M^{(x)}\}$, $\tr[O^{(x)}\Delta^{(y)}]=\delta_{xy}$. For an overcomplete IC instrument, choose a subfamily of its Choi operators that forms a basis of $W$ and use the corresponding dual basis. Biorthogonality forces linear independence, and testing against the spanning instrument subfamily shows that the block expansion is preserved. The common label set $x$ is thereby reduced to at most $\dim W$ elements. For $n\ge\ell$, split $[1,n]=M\cup F$, $M=[1,\ell]$, and apply \eqref{eq:blockexpansion} at the tripartition $P\mid M\mid F$. Contracting the past slots with $\xi_P^T$ and the future slots with $(\xi_n\otimes\cdots\otimes\xi_1)^T$ gives the displayed expansion, which exhibits $\Phi_{\xi_P}$ as an element of $\spn\{g_x\}$, a space of dimension at most the number of dual-set elements, $\le\dim W=d^{4\ell}$ \cite[Sec.~IV]{Taranto2019structure}. For each fixed $x$, the functionals $g_x$ obtained from different future-window lengths are the same object under identity insertion: appending identities to a future window leaves its value unchanged, and Definition~\ref{def:stationary}(i), together with uniqueness in the GET, makes this identification consistent across all windows. The expansion coefficients are therefore independent of $n$. The count $d^{4\ell}$ is precisely the dimension of the memory-block operator space in \cite[Sec.~IV]{Taranto2019structure}: each of the $\ell$ input/output pairs contributes a $d^4$-dimensional operator space. It is not the cardinality of an overcomplete instrument. Finally, the same block expansion applies to the stationary infinite past supplied by the GET. After contracting that past, every conditional future functional, including $\Phi^{\mathrm{stat}}_\varnothing$, is a linear combination of at most $\dim W$ of the $g_x$. No infinite sum is required. Hence $\Phi^{\mathrm{stat}}_\varnothing\in\cQ$, so its use as an order unit is legitimate. Windows shorter than $\ell$ are covered by marginalization. \qed

\subsection{The update map and the boundary state}

This subsection states and proves the two lemmas that the main text summarizes in prose: well-definedness of the transfer map (Lemma~\ref{lem:transfer}) and the boundary state with the generating formula (Lemma~\ref{lem:boundary}).

\begin{lemma}[Transfer map]\label{lem:transfer}
For $\xi\in\cV$, $(\bbE_{\hat\xi}\Phi)(\xi_1,\dots,\xi_n):=\Phi(\xi,\xi_1,\dots,\xi_n)$ defines a linear map $\bbE_{\hat\xi}:\cQ\to\cQ$ which depends only on $\Phi$ (not on any past generating it, nor on any instrument choice). $\bbE_{\hat\xi}\Phi_{\xi_P}=\Phi_{\xi_P\otimes\xi}$. $\xi\mapsto\bbE_{\hat\xi}$ is linear. The composition law $\bbE_{\hat\xi'}\circ\bbE_{\hat\xi}\,\Phi_{\xi_P}=\Phi_{\xi_P\otimes\xi\otimes\xi'}$ respects time order.
\end{lemma}

\begin{proof}
The definition uses only the values of $\Phi$ as a functional: if $\Phi_{\xi_P}=\Phi_{\xi'_P}$ as functionals (equal on all tuples of all lengths), their images are equal. This is the point at which the construction is intrinsic: no choice of past, instrument, or recovery map enters. The update rule follows by evaluating $(\bbE_{\hat\xi}\Phi_{\xi_P})(\xi_1,\dots,\xi_n)=\Phi_{\xi_P}(\xi,\xi_1,\dots,\xi_n)=\Phi_{\xi_P\otimes\xi}(\xi_1,\dots,\xi_n)$. Linearity in $\xi$ follows from multilinearity of the comb, and the composition law follows by applying the same evaluation identity twice, with the latest intervention placed to the right in the past block.
\end{proof}

\begin{lemma}[Boundary state and generating formula]\label{lem:boundary}
With $\rho(\Phi):=\Phi(\varnothing)$ and $\bbone_{\cQ}:=\Phi^{\mathrm{stat}}_\varnothing$ (the conditional future functional of the stationary infinite past, which exists by Proposition~\ref{prop:extension}), the generating formula \eqref{eq:generating} holds and $\rho\circ\bbE_{\cI}=\rho$.
\end{lemma}

\begin{proof}
Iterating the update rule of Lemma~\ref{lem:transfer} from the stationary infinite past gives $\bbE_{\hat\xi_n}\circ\cdots\circ\bbE_{\hat\xi_1}(\bbone_{\cQ})=\Phi^{\mathrm{stat}}_{(\xi_1,\dots,\xi_n)}$, the conditional future functional of the stationary past followed by the window $(\xi_1,\dots,\xi_n)$. Evaluating at the empty future,
\begin{equation}
\rho\big(\bbE_{\hat\xi_n}\circ\cdots\circ\bbE_{\hat\xi_1}(\bbone_{\cQ})\big)=\Phi^{\mathrm{stat}}_{(\xi_1,\dots,\xi_n)}(\varnothing)=\cT_{[1,n]}[\xi_1\otimes\cdots\otimes\xi_n],
\end{equation}
where the last equality is the definition of the stationary functional: the statistics of a window preceded by the stationary past are the stationary window statistics. This is \eqref{eq:generating}. The identity $\rho\circ\bbE_{\cI}=\rho$ is the special case $\hat{\cM}=\cI$ of the normalization identity proved in the identity-insertion unitality paragraph below.
\end{proof}

\subsection{Proof of Lemma~\ref{lem:order} (order structure)}

This subsection gives the full proof of Lemma~\ref{lem:order}. The main text presents only the key estimate, the ratio lemma of Eq.~\eqref{eq:ratio}.

\noindent\textbf{The two operational cones.} Let $\mathfrak{T}_{\mathrm{op}}$ consist of positive combinations of intervention tuples whose every slot is either a causal-break element $\varrho^{(b)}\otimes\Pi^{(a)}$ (with $\{\Pi^{(a)}\}$ a full-rank IC POVM and $\{\varrho^{(b)}\}$ an IC set of states) or the identity intervention $\cI$. The intervention cone $\mathcal O\subset\cV$ is the corresponding measure-and-prepare cone of positive Choi elements. The distinct functional cone $\cQ_+\subset\cQ$ is defined by $\Phi\ge0$ iff $\Phi(\xi)\ge0$ for all $\xi\in\mathfrak{T}_{\mathrm{op}}$. Matrix levels are defined entrywise. Since the Choi operator of the process is positive, every $\Phi_{\xi_P}$ with $\xi_P\ge0$ is positive, and $\bbone_{\cQ}\ge0$.

\noindent\textbf{Positivity for general positive $\xi$: what holds and what does not.} For a functional $\Phi_{\xi_P}$ generated by a single positive past ($\xi_P\ge0$) one has $\Phi_{\xi_P}(\xi_1,\dots,\xi_n)=\tr[\Upsilon(\xi_P\otimes\xi_1\otimes\cdots)^T]\ge0$ for every tuple of positive Choi elements, because both factors are positive operators. Positive reachable functionals therefore stay positive under $\bbE_{\hat\xi}$ for arbitrary positive $\xi$. However, a general element of $\cQ_+$ is a linear combination of such functionals, and the conclusion does not propagate to combinations: comb positivity alone does not control $\Phi(\xi\otimes\tau)$ for $\xi\otimes\tau\notin\mathfrak{T}_{\mathrm{op}}$. The CP statements of this paper are therefore asserted only for $\xi$ in $\mathcal O$. The general extension is Conjecture~\ref{conj:coherent}. What the ratio lemma adds is not positivity but domination by the order unit.

\noindent\textbf{Support alignment.} If $\xi_F\ge0$ satisfies $\bbone_{\cQ}(\xi_F)=\tr[\Upsilon^{\mathrm{stat}}_F\xi_F^T]=0$, then $\xi_F$ is supported outside $\operatorname{supp}\Upsilon^{\mathrm{stat}}_F$. A positive operator is supported within the tensor product of the supports of its marginals (its range is contained in theirs), so the joint comb $\Upsilon_{PF}$ (whose $F$-marginal is the stationary one) gives $\Phi_{\xi_P}(\xi_F)=\tr[\Upsilon_{PF}(\xi_P\otimes\xi_F)^T]=0$ for every past.

\noindent\textbf{The evaluation norm.} CPTP Choi operators linearly span $\cV$ (unitary channels and depolarizing combinations suffice). Hence a functional vanishing on all CPTP tuples of all lengths vanishes identically, so \eqref{eq:evalnorm} is a norm. $\|\bbone_{\cQ}\|_{\mathrm{eval}}=1$ by comb normalization. Since $\cI$ is a channel, prepending it maps CPTP tuples to CPTP tuples, so $\bbE_{\cI}$ is a contraction in this norm.

\noindent\textbf{The ratio lemma (order unit property).} For a break element $O^{(x_M)}$ on the first $\ell$ slots, the factorization \eqref{eq:markovorder} gives
\begin{equation}
\Phi_{\xi_P}\big(O^{(x_M)}\otimes\xi_{\mathrm{rest}}\big)=\beta_{x_M}(\xi_P)\,w_{x_M}(\xi_{\mathrm{rest}}),\qquad
\bbone_{\cQ}\big(O^{(x_M)}\otimes\xi_{\mathrm{rest}}\big)=p^{\mathrm{stat}}(x_M)\,w_{x_M}(\xi_{\mathrm{rest}}),
\end{equation}
with $\beta_{x_M}(\xi_P)=\tr[\Upsilon_P^{(x_M)}\xi_P^T]\ge0$ and $w_{x_M}(\xi_{\mathrm{rest}})=\tr[\Upsilon_F^{(x_M)}\xi_{\mathrm{rest}}^T]\ge0$ (partial inner products of positive operators). Full rank of the POVM gives the state-independent lower bound $p^{\mathrm{stat}}(x_M)\ge(\min_a\lambda_{\min}(\Pi^{(a)}))^\ell>0$: at each of the $\ell$ steps, the measurement outcome probability is bounded below by $\min_a\lambda_{\min}(\Pi^{(a)})$ regardless of the state being measured.

The history $\xi_P$ is allowed to be an arbitrary CP intervention tuple, with no trace-preserving, subnormalization, or probability normalization assumption. For each fixed $\xi_P$, every numerator $\beta_{x_M}(\xi_P)$ is finite and every stationary denominator is strictly positive. Therefore
\begin{equation}
0\le\Phi_{\xi_P}\le R(\xi_P)\,\bbone_{\cQ}\quad\text{on full-break tuples},\qquad
R(\xi_P):=\max_{x_M}\frac{\beta_{x_M}(\xi_P)}{p^{\mathrm{stat}}(x_M)}<\infty .
\end{equation}
This is an elementwise constant: $\xi_P\mapsto c\xi_P$ gives $R(\xi_P)\mapsto cR(\xi_P)$ for $c>0$. For fixed $\xi_P$ it is uniform over all future break tests and does not depend on the past length. Appending identity slots to the past leaves the functional and the ratio unchanged by identity-insertion consistency. No tester-probability interpretation of an unnormalized history, and no single bound uniform over all $\xi_P$, is used. The lower side of the two-sided domination is simply positivity of $\Phi_{\xi_P}$.

Extension to positive combinations is termwise by linearity. Identity absorption gives a uniform reduction for arbitrary interleavings: dropping identity slots, whether at the beginning, in the interior, or at the end, reduces a test tuple to a consecutive break-only tuple and changes neither the functional values nor the domination constants. If the reduced tuple contains at least $\ell$ breaks, the full-break ratio applies to its first $\ell$ slots and the remaining slots form $\xi_{\mathrm{rest}}$. If it contains $n<\ell$ breaks, it is one of finitely many partial-break patterns, and inserting identities again covers every original interleaving of the same reduced pattern. A tuple containing $n<\ell$ break elements is therefore bounded directly: $\Phi_{\xi_P}(O_1\otimes\cdots\otimes O_n)\le\rho(\Phi_{\xi_P})$ by positivity, while $\bbone_{\cQ}(O_1\otimes\cdots\otimes O_n)\ge(\min_a\lambda_{\min}(\Pi^{(a)}))^{n}\ge(\min_a\lambda_{\min}(\Pi^{(a)}))^{\ell}>0$. Each partial-break pattern therefore has a finite ratio, with the same scaling in $\xi_P$. Taking $\widetilde R(\xi_P)$ to be the maximum of $R(\xi_P)$ and these finitely many partial-break ratios covers the whole operational test set. Hence for $\Psi=\sum_j a_j\Phi_{\xi_{P_j}}$,
\begin{equation}
|\Psi(\xi)|\le t(\Psi)\,\bbone_{\cQ}(\xi)\quad\text{for all }\xi\in\mathfrak{T}_{\mathrm{op}},\qquad t(\Psi):=\textstyle\sum_j|a_j|\,\widetilde R(\xi_{P_j}),
\end{equation}
which is precisely the elementwise two-sided domination making $\bbone_{\cQ}$ an order unit for the operational cone. \qed

\noindent\textbf{Identity-insertion unitality and normalization.} $\bbE_{\cI}(\bbone_{\cQ})=\bbone_{\cQ}$ because identity insertion marginalizes and the stationary marginal of the stationary process is itself. Positivity of $\rho$ follows from $\rho(\Phi)=\Phi(\varnothing)$ with the empty-window value equal to the value at any single identity insertion. $\rho(\bbone_{\cQ})=1$. For any channel $\hat{\cM}$,
\begin{equation}
(\rho\circ\bbE_{\hat{\cM}})(\Phi_{\xi_P})=\Phi_{\xi_P}(\cM)=\cT[\xi_P\otimes\cM]=\cT[\xi_P]=\rho(\Phi_{\xi_P}),
\end{equation}
where the third equality is comb causality: a deterministic operation at the latest slot does not change earlier statistics (the Choi-level identity $\tr_{\mathtt{o}}\cM=\bbone_{\mathtt{i}}$ combined with the causality hierarchy). \qed

\subsection{Anchor cases}
\label{app:anchors}

This subsection works out the two anchor cases that calibrate the generating triple of Proposition~\ref{prop:triple} in the main text.

\noindent\textbf{Classical diagonal processes.} If all intervention operators are diagonal in a fixed basis, the process is a classical stationary process. Finite Markov order $\ell$ relative to the break is a classical order-$\ell$ Markov chain. Then $\cQ$ is the function space on the last $\ell$ symbols ($\dim=|\mathcal{X}|^\ell$), $\bbE_{\cI}$ is the shift, $\rho$ the stationary measure, and the ratio lemma is the classical bound $|\Psi/\bbone|\le\max\mu/\pi\le1/\pi_{\min}$.

\noindent\textbf{Markov processes.} For $\ell=1$ with IC preparations, $\cQ$ is (dual to) the system state space, $\bbE_{\hat{\cM}}$ is the predual action of $\hat{\cM}$, $\bbE_{\cI}=\mathrm{id}$, and normalization is trace preservation.

\section{Details for Sec.~\ref{sec:spectral}: proofs of the spectral criteria}
\label{app:spectral}

This appendix proves the spectral criteria and the record ergodic theorem of Sec.~\ref{sec:spectral}: Theorems \ref{thm:ergodic}, \ref{thm:gap}, \ref{thm:converse}, and \ref{thm:record}, together with the base-norm criterion (Lemma~\ref{lem:basenorm}) and the classical limit theorems (Lemma~\ref{lem:classical-clt}) that the record theorem uses.

\noindent\textbf{Power-boundedness and the mean ergodic projection.} $\bbE_{\cI}$ is a contraction in the evaluation norm (Lemma~\ref{lem:order}), hence power-bounded. On a finite-dimensional space, power-boundedness implies that all Jordan blocks at peripheral eigenvalues are trivial and that the Ces\`aro averages $\frac1N\sum_{\tau<N}\bbE_{\cI}^\tau$ converge to the projection $P$ onto the fixed-point space along $\operatorname{ran}(\mathrm{id}-\bbE_{\cI})$.

\noindent\textbf{Proof of Theorem~\ref{thm:ergodic} (ergodicity criterion).} (i)$\Rightarrow$(iii): if the fixed-point space is one-dimensional, $P=\bbone\otimes\varphi$ for an invariant functional $\varphi$. Invariance $\rho\circ\bbE_{\cI}=\rho$ and the one-dimensionality of the invariant functionals (which coincides with the one-dimensionality of the fixed points, since power-boundedness makes the algebraic and geometric multiplicities of the eigenvalue $1$ coincide) give $\varphi=\rho$. (iii)$\Rightarrow$(ii): the Ces\`aro average of \eqref{eq:idlegap} inserts $P$ between $\bbE_{\hat\xi_B}$ and $\bbE_{\hat\xi_A}(\bbone)$, and the rank-one form makes it vanish. (ii)$\Rightarrow$(i): a second fixed point $\Phi_0$ would, by right spanning, be represented exactly as a linear combination of $\bbE_{\hat\xi_A}(\bbone)$ over finitely many blocks (the span is all of $\cQ$, not merely dense), producing a nondecaying Ces\`aro correlation. \qed

\noindent\textbf{Proof of Theorem~\ref{thm:gap} (gap implies mixing).} With trivial peripheral spectrum, $\bbE_{\cI}=P+R$, $PR=RP=0$, $\spec R=\spec\bbE_{\cI}\setminus\{1\}$, and for every $\lambda>|\nu_2|$ one has $\|R^\tau\|\le C\lambda^\tau$ (Jordan blocks at the subleading eigenvalues contribute polynomial factors, absorbed by the strict inequality). Substituting into \eqref{eq:idlegap} gives the bound with $K_{\xi_A,\xi_B}=C\,\|\rho\circ\bbE_{\hat\xi_B}\|\,\|\bbE_{\hat\xi_A}(\bbone)\|$. \qed

\noindent\textbf{Proof of Theorem~\ref{thm:converse} (mixing implies gap).} By two-sided spanning, pointwise decay of \eqref{eq:idlegap} for all blocks implies $\bbE_{\cI}^\tau\to\bbone\otimes\rho$ strongly (finite-dimensional). Any peripheral eigenvalue $e^{i\theta}\ne1$ with eigenvector $u$ would contribute a nondecaying oscillation $e^{i\theta\tau}$ to some correlation (choose $\xi_A$ with $\bbE_{\hat\xi_A}(\bbone)$ overlapping $u$ and $\xi_B$ with $\rho\circ\bbE_{\hat\xi_B}$ nonvanishing on $u$), contradiction. \qed

\noindent\textbf{Proof of Theorem~\ref{thm:record} (exponential mixing of records).} \emph{Event algebra and stationarity.} The record is a sequence over the finite outcome alphabet of $\cJ$. Events are cylinder events (finite unions of outcome strings on finite intervals). The record family is stationary: shift-invariance of $p(x_1,\dots,x_m)=T_{[1,m]}[\cM_{x_1}\otimes\cdots\otimes\cM_{x_m}]$ is the translation covariance of the comb family, and tail consistency follows from the normalization condition (summing the latest outcome applies a channel at the latest slot, which changes no earlier statistics). One-sided Kolmogorov consistency requires no more. It suffices to control $P(A\cap B)-P(A)P(B)$ for $A$ on $[1,m]$ and $B$ on $[m+n+1,m+n+m']$ with constants independent of $m$, $m'$, and position.

\emph{The uniform event bound.} For an outcome string $s=(i_1,\dots,i_m)$ write $\bbE_s:=\bbE_{i_m}\circ\cdots\circ\bbE_{i_1}$, and for an event $A$ set $\Gamma_A:=\sum_{s\in A}\bbE_s$. We work in the \emph{base norm} of the operational order on the self-adjoint part of $\cQ$,
\begin{equation}
\|\Psi\|_1:=\inf\{\rho(\Psi_+)+\rho(\Psi_-)\;\big|\;\Psi=\Psi_+-\Psi_-,\ \Psi_\pm\ge0\},
\end{equation}
a genuine norm because the operational cone is proper and generating \cite{Paulsen2002}, and because $\rho$ is strictly positive on it. For the latter, work on the cone generated by reachable positive functionals. If a generator $G=\Phi_{\xi_P}$ satisfies $\rho(G)=0$, then on every joint past/future window the positive operator $G_F:=\tr_P[\Upsilon_{PF}\xi_P^T]$ has trace $\rho(G)=0$ and hence vanishes. Consequently $G(\xi_F)=\tr[G_F\xi_F^T]=0$ for every future intervention, so the spanning property of the operational tests gives $G\equiv0$. A general cone element is a positive combination $\sum_j a_j G_j$ of generators, and $\rho(\Phi)=\sum_j a_j\,\rho(G_j)=0$ with all terms nonnegative forces each summand to vanish. The record-relevant elements lie in the same operational order. Each $\bbE_{\cJ}^{\,n}\bbone$ is a reachable positive functional, so $e_{\cJ}=\lim_n\bbE_{\cJ}^{\,n}\bbone$ lies in the closure of that cone. Since $\cQ_+$ is closed in the finite-dimensional space $\cQ$ (Remark~\ref{rem:opcone}), this closed cone is contained in $\cQ_+$. In particular, $e_{\cJ}\in\cQ_+$. Likewise $\Gamma_A(\bbone)$ is a reachable positive functional. The zero-marginal argument extends by continuity to this closure and to all positive outcome images, so $\rho$ is strictly positive on every nonzero element of the record-relevant cone containing $\Gamma_A(\bbone)$ and $e_{\cJ}$. Positivity of the outcome maps gives $0\le\Gamma_A\Psi\le\bbE_{\cJ}^{\,m}\Psi$ pointwise on the test set for positive $\Psi$. The normalization identity $\rho\circ\bbE_{\cJ}=\rho$ (the instrument sums to a channel) is what makes the bound uniform: for positive $\Psi$,
\begin{equation}
\|\Gamma_A\Psi\|_1=\rho(\Gamma_A\Psi)\le\rho(\bbE_{\cJ}^{\,m}\Psi)=\rho(\Psi)=\|\Psi\|_1,
\end{equation}
so $\|\Gamma_A\|_1\le1$ for every event of every length. The readout side obeys the same bound, $|\rho(\Gamma_B\Psi)|\le\|\Psi\|_1$, and $\|\Gamma_A(\bbone)\|_1=\rho(\Gamma_A(\bbone))=P(A)\le1$. (The order-unit norm would not give this uniformity directly, since $\bbE_{\cJ}$ is not unital. The base norm, with $\rho\circ\bbE_{\cJ}=\rho$, is the right tool.)

\emph{Mixing estimate.} With $\bbE_{\cJ}^{\,n}=e_{\cJ}\otimes\rho+R^n$, the peripheral hypothesis gives $\|R^n\|_1\le C\lambda^n$ (finite-dimensionality. Possible Jordan-block polynomial factors are absorbed into $C$ at the cost of an arbitrarily small increase of $\lambda$). Then
\begin{equation}
\big|P(A\cap B)-P(A)P(B)\big|=\big|\rho\big(\Gamma_B R^n\Gamma_A(\bbone)\big)\big|\le C\lambda^n,
\end{equation}
with $C$ independent of the events, their interval lengths, and their positions, where $P(B)=\rho(\Gamma_B e_{\cJ})$ is the stationary one-sided marginal. Indeed, the marginal of $B$ at position $n$ is $\rho(\Gamma_B\bbE_{\cJ}^{\,n}\bbone)$. It is independent of $n$ by stationarity of the record and converges to $\rho(\Gamma_B e_{\cJ})\,\rho(\bbone)=\rho(\Gamma_B e_{\cJ})$ by the spectral hypothesis. Combining these two facts gives, for every cylinder event $B$,
\begin{equation}
\rho(\Gamma_B e_{\cJ})=\rho(\Gamma_B\bbone)=P(B),\qquad \rho\circ\Gamma_B(\bbone-e_{\cJ})=0 .
\end{equation}
This is exponential $\alpha$-mixing, $\alpha(n)\le K\lambda^n$.

\emph{Limit theorems.} The probability model is a stationary process on a finite outcome alphabet, and $f$ is bounded, so every required moment is finite. Exponential mixing gives $\sum_n\alpha(n)\le K\sum_n\lambda^n<\infty$. When applied to the centered record variable, the covariance estimate gives $|\gamma_k|\le C'\lambda^k$, hence $\sum_k|\gamma_k|<\infty$ and the long-run variance $\sigma^2=\gamma_0+2\sum_{k\ge1}\gamma_k$ converges absolutely. These are, respectively, the stationarity, moment, mixing-summability, and covariance-summability inputs to the classical LLN and CLT for stationary $\alpha$-mixing sequences, see Lemma~\ref{lem:classical-clt} below. If $\sigma^2=0$ the CLT is degenerate in the precise sense that $(S_N-N\mu)/\sqrt N\to0$ in probability. \qed

\subsection{The base-norm criterion}

The base norm used in the proof of Theorem~\ref{thm:record} above is controlled by the following criterion, which is the content that makes the event bound uniform.

\begin{lemma}[Base-norm criterion]\label{lem:basenorm}
Let $K\subset V$ be a proper generating cone in a finite-dimensional space, $\rho\in K^*$, and define on $V_{\mathrm{sa}}=K-K$ the base seminorm $\|\Psi\|_1:=\inf\{\rho(\Psi_+)+\rho(\Psi_-)\mid \Psi=\Psi_+-\Psi_-,\ \Psi_\pm\in K\}$. Then $\|\Psi\|_1=\rho(\Psi)$ for $\Psi\in K$, and $\|\cdot\|_1$ is a norm if and only if $\rho$ is strictly positive on $K\setminus\{0\}$. In the setting of this paper, strict positivity holds on the cone generated by reachable positive functionals: if $G=\Phi_{\xi_P}$ and $\rho(G)=0$, the zero-marginal argument above gives $\tr_P[\Upsilon_{PF}\xi_P^T]=0$ on every future window, hence $G$ vanishes on every future intervention. The conclusion extends to positive combinations since $\rho$ is additive and nonnegative on each generator.
\end{lemma}

\begin{proof}
For $\Psi\in K$, the decomposition $\Psi=\Psi-0$ gives $\|\Psi\|_1\le\rho(\Psi)$, while any decomposition satisfies $\rho(\Psi)=\rho(\Psi_+)-\rho(\Psi_-)\le\rho(\Psi_+)+\rho(\Psi_-)$, so $\rho(\Psi)\le\|\Psi\|_1$. Equality follows. The seminorm axioms (homogeneity and the triangle inequality) hold by taking infima over decompositions. If $\rho(x)=0$ for some $x\in K\setminus\{0\}$, then $\|x\|_1=\rho(x)=0$ and the seminorm is not definite. Conversely, if $\rho>0$ on $K\setminus\{0\}$, then with $S=\{x\in K:|x|=1\}$ compact (any Euclidean norm $|\cdot|$), $c:=\min_{x\in S}\rho(x)>0$ by continuity, so $\rho(x)\ge c|x|$ on $K$, and every admissible decomposition gives $\rho(\Psi_+)+\rho(\Psi_-)\ge c(|\Psi_+|+|\Psi_-|)\ge c|\Psi|$. Hence $\|\Psi\|_1\ge c|\Psi|$ and the seminorm is definite. The final claim is the zero-marginal argument recalled in the statement. (That proper and generating alone do not suffice can be seen on $V=\mathbb{R}^2$, $K=\mathbb{R}_+^2$, $\rho(x,y)=x$.)
\end{proof}

\subsection{The classical limit theorems used in Theorem~\ref{thm:record}}

For completeness we record the classical input in the form in which it is used, with proof.

\begin{lemma}[LLN and CLT for the record]\label{lem:classical-clt}
Let $\{Y_t\}_{t\in\mathbb{Z}}$ be a stationary process on a finite alphabet whose cylinder events satisfy $|P(A\cap B)-P(A)P(B)|\le C\lambda^n$ whenever $A$ is measurable on the coordinates $\le0$ and $B$ on those $\ge n$. Then $\alpha(n)\le C\lambda^n$ for the strong mixing coefficients. If $X_t=f(Y_t)$ for a bounded real-valued function $f$, $S_N=\sum_{t=1}^N X_t$, $\mu=\mathbb{E}X_0$, then $S_N/N\to\mu$ almost surely, the series $\sigma^2=\operatorname{Var}(X_0)+2\sum_{k\ge1}\operatorname{Cov}(X_0,X_k)$ converges absolutely, and $(S_N-N\mu)/\sqrt{N}\Rightarrow N(0,\sigma^2)$, with a degenerate limit if $\sigma^2=0$. The standardized form requires $\sigma^2>0$.
\end{lemma}

\begin{proof}
Write $Z_t=X_t-\mu$, $T_N=\sum_{t\le N}Z_t$, and let $D$ bound $|Z_t|$.

\emph{Step 1: extension to the tail $\sigma$-algebras.} Fix $n$ and write $\mathcal F=\sigma(Y_t:t\le0)$, $\mathcal G_n=\sigma(Y_t:t\ge n)$, $c_n=C\lambda^n$. For a past cylinder event $A$, the class $\mathcal D_A=\{B\in\mathcal G_n:|P(A\cap B)-P(A)P(B)|\le c_n\}$ contains the future cylinders and is a monotone class by continuity of probability, hence equals $\mathcal G_n$ by the monotone class theorem. The symmetric argument in $A$ gives the estimate on $\mathcal F\times\mathcal G_n$, which is $\alpha(n)\le C\lambda^n$.

\emph{Step 2: a covariance bound.} For $U,V$ measurable on the past and future respectively with $|U|\le a$, $|V|\le b$, one has $|\operatorname{Cov}(U,V)|\le4ab\,\alpha(n)$ (write $W=\mathbb E[V\mid\mathcal F]-\mathbb EV$, estimate $\mathbb E|W|$ via the layer-cake representation and the mixing definition). Hence $|\gamma(k)|\le4D^2C\lambda^k$ for $k\ge1$, where $\gamma(k)=\mathbb E[Z_0Z_k]$. The covariance series converges absolutely, with $H:=|\gamma(0)|+2\sum_{k\ge1}|\gamma(k)|\le D^2+8D^2C\lambda/(1-\lambda)$.

\emph{Step 3: variance asymptotics and the LLN.} Stationarity gives $\operatorname{Var}T_N=N\gamma(0)+2\sum_{k<N}(N-k)\gamma(k)$, so $\operatorname{Var}T_N/N\to\sigma^2$ (the difference is bounded by $\frac2N\sum_{k<N}k|\gamma(k)|+2\sum_{k\ge N}|\gamma(k)|\to0$). Chebyshev along the subsequence $N_m=m^2$ gives a summable deviation bound, hence $T_{m^2}/m^2\to0$ a.s.. The sandwich $m^2\le N<(m+1)^2$ extends this to all $N$.

\emph{Step 4: the CLT.} Use the big-block/small-block decomposition with $p_N=\lfloor N^{1/3}\rfloor$, $q_N=\lfloor N^{1/4}\rfloor$: the big blocks are identically distributed by stationarity, their number satisfies $k_Np_N/N\to1$, the small blocks contribute $o(\sqrt N)$ in $L^2$ by Step 2, and the block dependence is controlled by $k_N\lambda^{q_N}\to0$ (comparing the characteristic function of the block sum with the product of the marginal characteristic functions via repeated use of Step 2). The marginal triangular array satisfies the Lindeberg condition trivially (the normalized blocks are bounded by $Dp_N/\sqrt N\to0$), so its characteristic function converges to $e^{-\sigma^2t^2/2}$. Pointwise convergence of characteristic functions to a function continuous at $0$ gives the weak convergence. If $\sigma^2=0$, Step 3 gives $T_N/\sqrt N\to0$ in probability directly.
\end{proof}

\section{Details for Sec.~\ref{sec:discussion}: a classical key-locked counterexample (self-contained)}
\label{app:keylock}

This appendix supplies a self-contained version of the classical permutation-gate counterexample cited in Sec.~\ref{sec:discussion}, in which the process mixes while the record does not. The construction is short, and we include it for self-containedness. Let $U$ be the permutation gate on two qubits $|00\rangle\mapsto|00\rangle$, $|01\rangle\mapsto|10\rangle$, $|10\rangle\mapsto|11\rangle$, $|11\rangle\mapsto|01\rangle$ (a 3-cycle plus a fixed point), inserted in the two-collision step of the collision model of Sec.~\ref{sec:numerics}. On the idle side, one checks on the computational basis that the joint step map obeys $N^2(X)=\tr(X)\,|00\rangle\langle00|$ for every operator $X$, so $\spec(N)=\{1,0,\dots,0\}$: the idle process mixes in two steps. On the driven side, with the break instrument ``measure $\sigma_z$, reprepare $|0\rangle$'', the classical ancilla bit is preserved forever: $U|0,i\rangle=|i,0\rangle$ writes the memory bit into the system and $U|i,0\rangle=|i,i\rangle$ writes it into the fresh ancilla, so the driven memory channel has spectrum $\{1,1,0,0\}$, and the record is a constant sequence determined by the initial memory bit. In the symmetric mixture of the two regimes the record autocorrelation is identically $1$: the process mixes while the record is non-ergodic. The preceding computation is self-contained and establishes the counterexample used in this paper.

\section{Details for Sec.~\ref{sec:numerics}: the collision-model study}
\label{app:numerics}

This appendix supplies the closed-form formulas, sanity checks, and rate-versus-spectrum comparisons behind the collision-model study of Sec.~\ref{sec:numerics}.

\subsection{Exact formulas}

The step structure is: instrument on $S$. Collision $U$ on $(S,a_{\mathrm{old}})$, discard $a_{\mathrm{old}}$. Collision $U$ on $(S,a_{\mathrm{new}})$, $a_{\mathrm{new}}=|0\rangle\langle0|$. With reprepare state $\sigma_{\mathrm{prep}}$, define
\begin{equation}
\mathrm{joint}(\eta):=U\Big(\tr_b\big[U(\sigma_{\mathrm{prep}}\otimes\eta)U^\dagger\big]\otimes|0\rangle\langle0|\Big)U^\dagger,\qquad
M(\eta)=\tr_S\,\mathrm{joint}(\eta),
\end{equation}
and the outcome-selective maps $\widetilde M_x(\eta)=(\langle x|\otimes I_c)\,\mathrm{joint}(\eta)\,(|x\rangle\otimes I_c)$, satisfying $\sum_x\widetilde M_x=M$ exactly. The record probabilities obey the chain rule $p(x_{k+1},\dots,x_{k+n}\,|\,\eta_k)=\tr\big[\widetilde M_{x_{k+n}}\circ M^{\,n-1}\circ\widetilde M_{x_{k+1}}(\eta_k)\big]$, and the ACF is
\begin{equation}
C(\tau)=\sum_{x,y}s_xs_y\,\tr\big[\widetilde M_y\,M^{\tau-1}\,\widetilde M_x\,\eta_*\big]-\mathbb{E}[s_t]^2,\qquad s_k=2x_k-1,
\end{equation}
with $\eta_*$ the fixed point of $M$. The idle step map on the joint memory $(S,a)$ is
\begin{equation}
N(\rho)=U\Big(\tr_a\big[U\rho\,U^\dagger\big]\otimes|0\rangle\langle0|\Big)U^\dagger .
\end{equation}
For the base gate, $N=Q\circ P$ factors through a four-dimensional image, and its nonzero spectrum equals that of the Pauli-diagonal channel $T(\sigma)=\cos^2(2\theta)\,\sigma+\sin^2(2\theta)\,X\sigma X$, namely $\{1,1,\cos4\theta,\cos4\theta\}$. The second eigenvalue $1$ reflects the conserved functional $\tr[(X_S\otimes I)\,\cdot\,]$. The $z$-preparation/$z$-measurement idle-gap correlation has the closed form $\Delta(\tau)=-\cos^2(2\theta)(\cos4\theta)^{\tau-1}$. For the extended gate, the nonzero spectrum of $N$ is $\{1,\cos2\theta,\cos4\theta,\cos2\theta\cos4\theta\}$ (numerically matched to $\le4\times10^{-15}$ at all tested $\theta$).

\subsection{Sanity checks}

Trace preservation of $M$ and $N$ to $\sim10^{-16}$. $\sum_x\widetilde M_x=M$ exactly. $\eta_*$ a valid density matrix with Bloch vector $(0,-\sin2\theta,\cos2\theta)$. Record probabilities normalized to $1-10^{-16}$. Eigenvalue $1$ present in both spectra to $\le6\times10^{-16}$. The i.i.d.\ limit recovered as $\theta\to0$ ($|C(1)|\sim10^{-6}$ at $\theta=10^{-3}$, $|C(\tau\ge2)|\sim10^{-16}$). All matrix algebra is dense and in IEEE double precision. At the period-$2$ point $\theta=\pi/4$, and wherever eigenvalue branches cross, the spectra are evaluated from the closed forms of this section. Numerical diagonalization is a cross-check, which is the content of the matching bounds quoted above.

\subsection{Rates versus spectra}

Record level (extended gate): fitted ACF decay rates equal $|\nu_2(M)|$ to six significant digits ($\theta=0.2,0.4,0.6,0.8$. See the table in Sec.~\ref{sec:numerics}). Idle level (base gate): the closed form matches numerics to $\le3\times10^{-15}$. Cross-level: $\nu_2(M)\le\nu_2(N)$ at all tested points. Idle mixing implies record mixing at all tested points. The converse fails (base gate: record i.i.d., idle non-mixing). The rank-one phenomenon of the base gate ($\nu_2(M)=0$ for every measure-and-prepare instrument) is proved by writing the write map $G(A)=\tr_S[U(A\otimes|0\rangle\langle0|)U^\dagger]$ and read map $F(\eta)=\tr_b[U(\sigma_{\mathrm{prep}}\otimes\eta)U^\dagger]$ and observing that $G$ reads only $(\tr A,\tr(XA))$ while $\tr(XF(\eta))$ is independent of the variable part of $\eta$.

\section{Details for Sec.~\ref{sec:spectral}: proof of the record-invisibility theorem}
\label{app:nbl}

This appendix proves Theorem~\ref{thm:nbl}. The peripheral structure used here is the channel-level theorem of Ref.~\cite{WolfPG2010}. The operator-system-level analogue for the transfer map remains open (Sec.~\ref{sec:discussion}(iii)). The argument uses the peripheral eigenstructure of a finite-dimensional channel, recovery through a composition, and complementarity between correctable and private information. Throughout, channels act on density operators, and the peripheral algebra is understood with the matrix order and modified product supplied by the peripheral eigenstructure theorem. The faithful stationary state assumed in the theorem provides the reference state needed for both the peripheral projection and Petz recovery.

We first isolate the recovery statement used for the two-collision decomposition. Its conclusion concerns recovery on the intermediate image. That image need not be an algebra.

\begin{corollary}[Reversibility propagates through compositions]\label{cor:petz}
Let $\Phi_1:\mathcal A_0\to\mathcal A_1$ and $\Phi_2:\mathcal A_1\to\mathcal A_2$ be finite-dimensional quantum channels, and let $\mathcal B\subseteq\mathcal A_0$ be a subalgebra on which $\Phi_2\circ\Phi_1$ is reversible. Then $\Phi_1$ is reversible on $\mathcal B$, and $\Phi_2$ is reversible on $\Phi_1(\mathcal B)$ as a recovery relation.
\end{corollary}

\begin{proof}
For states $\rho,\sigma$ on $\mathcal B$ with finite relative entropy, reversibility of the composite is equivalent to equality in the data-processing inequality,
\begin{equation}
S(\Phi_2\Phi_1\rho\,\|\,\Phi_2\Phi_1\sigma)=S(\rho\,\|\,\sigma),
\end{equation}
by Petz's sufficiency theorem \cite{Petz1986,Petz1988,Jencova2024}. Data processing at the two factors gives
\begin{equation}
S(\rho\,\|\,\sigma)
\ge S(\Phi_1\rho\,\|\,\Phi_1\sigma)
\ge S(\Phi_2\Phi_1\rho\,\|\,\Phi_2\Phi_1\sigma)
=S(\rho\,\|\,\sigma).
\end{equation}
Both inequalities are therefore equalities for every such pair. The first family of equalities is sufficiency of $\Phi_1$ on $\mathcal B$, and the second is sufficiency of $\Phi_2$ on $\Phi_1(\mathcal B)$. More precisely, Petz's criterion with a fixed faithful reference state on $\mathcal B$ yields one recovery map for $\Phi_1$ valid on all of $\mathcal B$, and one recovery map for $\Phi_2$ valid on all of $\Phi_1(\mathcal B)$. This is a recovery relation on $\Phi_1(\mathcal B)$. It neither requires nor implies that this intermediate image is itself an algebra.
\end{proof}

\begin{proof}[Proof of Theorem~\ref{thm:nbl}]
\textbf{Step 1: peripheral blockwise reversibility.} Let $\mathfrak A_{\mathrm{per}}$ be the peripheral eigenspace of $\Phi$. By the Wolf--P\'erez-Garc\'ia peripheral eigenstructure theorem \cite[Thms.~8--9]{WolfPG2010}, $\mathfrak A_{\mathrm{per}}$ is a finite-dimensional C$^*$-algebra for the modified product, the peripheral projection is a conditional expectation onto it, and $\Phi$ acts on its simple blocks by unitary conjugation together with permutations of isomorphic blocks. By hypothesis, $\mathcal A$ is one invariant noncommutative simple summand, so no permutation moves it to another summand. The restriction is therefore a $*$-automorphism of $\mathcal A$, hence
\begin{equation}
\Phi(X)=VXV^\dagger\qquad(X\in\mathcal A)
\end{equation}
for a unitary $V$. On this invariant simple block, $\mathrm{Ad}_V$ is a $*$-automorphism for the ordinary matrix product, and the modified product therefore coincides with ordinary matrix multiplication. Hence the subsystem form of KKS complementarity applies without any further change of product. In particular, $\Phi$ is perfectly reversible on $\mathcal A$, with inverse $\mathrm{Ad}_{V^\dagger}$ there.

\textbf{Step 2: propagation through the collisions.} In the collision-model realization, $\Phi=\Phi_2\circ\Phi_1$. Corollary~\ref{cor:petz}, applied with $\mathcal B=\mathcal A$, shows that $\Phi_1$ is reversible on $\mathcal A$ and $\Phi_2$ is reversible on $\Phi_1(\mathcal A)$. Thus neither collision can destroy information whose composition is perfectly preserved. The intermediate image $\Phi_1(\mathcal A)$ need not be an algebra, so complementarity will not be applied to $\Phi_2$ on that image. The next step uses only the composed channel, on which $\mathcal A$ is a genuine peripheral algebra block.

\textbf{Step 3: complementarity and readout factorization.} Since $\mathrm{Ad}_{V^\dagger}$ is a recovery channel for $\Phi$ on $\mathcal A$, this simple block is a correctable subsystem for $\Phi$. The complementarity theorem of Kretschmann, Kribs, and Spekkens \cite[Thm.~5]{KKS2008} therefore gives a fixed environment state $\tau$ such that
\begin{equation}\label{eq:nbl-complement}
\Phi^c(X)=\tr(X)\,\tau\qquad(X\in\mathcal A).
\end{equation}
In an explicit Stinespring dilation of the instrument-generated channel, let $S_{\mathrm r}$ denote the measured-system output supplied by the readout collision, $E_{\mathrm c,1}$ and $E_{\mathrm c,2}$ the environments of the non-readout and readout collisions, respectively, and $E_{\mathrm I}$ the instrument factor, including its outcome and repreparation register. The complementary channel $\Phi^c$ therefore has output on $S_{\mathrm r}\otimes E_{\mathrm c,1}\otimes E_{\mathrm c,2}\otimes E_{\mathrm I}$. Collecting every factor other than the measured system into $E_1:=E_{\mathrm c,1}\otimes E_{\mathrm c,2}\otimes E_{\mathrm I}$, the fixed measured-system marginal is $\mathcal C=\tr_{E_1}\circ\Phi^c$. Thus $E_1$ in this shorthand is the joint instrument-and-collision environment complementary to $S_{\mathrm r}$, not just one collision environment. The same fixed-marginal definition applies to the general instrument-generated realization. Equation~\eqref{eq:nbl-complement} consequently yields
\begin{equation}
\mathcal C(X)=\tr(X)\,\sigma_{\mathcal C},
\qquad \sigma_{\mathcal C}:=\tr_{E_1}\tau,
\qquad X\in\mathcal A.
\end{equation}
For the fixed readout POVM $\{F_i\}$, set $p_i:=\tr(F_i\sigma_{\mathcal C})$. Then $p_i\ge0$, $\sum_i p_i=1$, and
\begin{equation}
r_i(X)=\tr[F_i\mathcal C(X)]=\tr(X)\,p_i,
\end{equation}
which is Eq.~\eqref{eq:nbl-readout}. Finally, $\Phi^n(X)\in\mathcal A$ and $\tr[\Phi^n(X)]=\tr(X)$ for every $n$, by Step 1. Iterating the same identity therefore shows that every finite-string record probability generated by the fixed instrument depends on $X\in\mathcal A$ only through $\tr(X)$. For a normalized state even this factor is $1$. Every bounded real-valued function of the record is correspondingly constant on the $\mathcal A$-content. Thus exact preservation of the noncommutative simple block makes that block invisible to the record. \qed
\end{proof}

The role of each hypothesis is visible in the proof. Complementarity applies within one correctable simple block, while a nontrivial center labels different blocks and can remain visible to the environment. If the peripheral algebra is commutative, its content is broadcastable and the trace-factorization conclusion need not hold. The classical key-locked example of Appendix~\ref{app:keylock} uses precisely this remaining freedom.